%% file: manuscript.tex
\documentclass[review]{elsarticle}

\usepackage{mathtools}
\usepackage{lineno}
\modulolinenumbers[5]
\usepackage{booktabs}
\usepackage{multirow}
\usepackage{graphicx}
\usepackage{boldline} 
\usepackage{array}
\usepackage{caption}
\usepackage[export]{adjustbox}
\usepackage[
            ]{hyperref}
\usepackage[acronym]{glossaries}
\usepackage[superscript,biblabel]{cite}
\usepackage{longtable}
\usepackage{tabularray}
\usepackage{xcolor}
\usepackage{float}
\usepackage[export]{adjustbox}
\usepackage{geometry}
\newgeometry{top=1.5in, bottom=1.5in,right=1.5in,left=1.5in}
\usepackage{gensymb}

\usepackage[resetlabels,labeled]{multibib}
\newcites{S}{Supplementary References}
\journal{Applied Energy}

\newcommand{\beginsupplement}{%
	\setcounter{table}{0}
	\renewcommand{\thetable}{S\arabic{table}}%
	\setcounter{figure}{0}
	\renewcommand{\thefigure}{S\arabic{figure}}%
	\setcounter{section}{0}
	\renewcommand{\thesection}{Supplemental \arabic{section}}%
	\setcounter{equation}{0}
	\renewcommand{\theequation}{S\arabic{equation}}
	\setcounter{page}{1}
}

\makeglossaries

\newacronym{ac}{AC}{Alternating current}
\newacronym{ccgt}{CCGT}{Closed cycle gas turbine}
\newacronym{chp}{CHP}{Combined heat and power}
\newacronym{dac}{DAC}{Direct air capture}
\newacronym{dc}{DC}{Direct current}
\newacronym{dpv}{DPV}{Distributed photovoltaics}
\newacronym{ev}{EV}{Electric vehicle}
\newacronym{hv}{HV}{High voltage}
\newacronym{lv}{LV}{Low voltage}
\newacronym{ocgt}{OCGT}{Open cycle gas turbine}
\newacronym{phs}{PHS}{Pumped hydro storage}
\newacronym{pv}{PV}{Photovoltaics}

\begin{document}

\begin{frontmatter}

\title{Distributed photovoltaics provides key benefits for a highly renewable European energy system
}

\author[mpe]{Parisa Rahdan\corref{cor1}  }
\cortext[cor1]{Lead contact and corresponding author, Email: parisr@mpe.au.dk}
\author[TUB]{Elisabeth Zeyen}   
\author[upm]{Cristobal Gallego-Castillo}
\author[mpe,novo]{Marta Victoria}

\affiliation[mpe]{organization={Department of Mechanical and Production Engineering and iCLIMATE Interdisciplinary Centre for Climate Change, Aarhus University},
            addressline={}, 
            postcode={8000}, 
            city={Aarhus},
            country={Denmark}}

\affiliation[TUB]{organization={Department of Digital Transformation in Energy Systems, Technische Universität Berlin Einsteinufer 25 (TA 8)},
            postcode={10587}, 
            city={Berlin},
            country={Germany}}     

\affiliation[upm]{organization={Aircraft and Space Vehicles Department, Universidad Politécnica de Madrid, Plaza Cardenal Cisneros 3},
             postcode={28040}, 
            city={Madrid},
            country={Spain}} 
            
\affiliation[novo]{organization={Novo Nordisk Foundation CO2 Research Center},
            addressline={Gustav Wieds Vej 10}, 
            postcode={8000}, 
            city={Aarhus},
            country={Denmark}}

 \begin{abstract}
 
Distributed solar photovoltaic (PV) systems are projected to be a key contributor to future energy landscape, but are often poorly represented in energy models due to their distributed nature. They have higher costs compared to utility PV, but offer additional advantages, e.g., in terms of social acceptance. Here, we model the European power network with a high spatial resolution of 181 nodes and a 2-hourly temporal resolution. We use a simplified model of distribution and transmission networks that allows the representation of power distribution losses and differentiates between utility and distributed generation and storage. Three scenarios, including a sector-coupled scenario with heating, transport, and industry are investigated. The results show that incorporating distributed solar PV leads to total system cost reduction in all scenarios (1.4\% for power sector, 1.9-3.7\% for sector-coupled). The achieved cost reductions primarily stem from demand peak reduction and lower distribution capacity requirements because of self-consumption from distributed solar. This also enhances self-sufficiency for countries. The role of distributed PV is noteworthy in the sector-coupled scenario and is helped by other distributed technologies including heat pumps and electric vehicle batteries.
\end{abstract}


\end{frontmatter}


\section{Introduction}

PV systems are expected to become a leading energy producer in many regions as they have very competitive costs that are expected to decrease even further due to technology learning \cite{victoria_2021, IEA_2020}. Several studies \cite{victoria_2021, haegel2023photovoltaics} have argued that neither material and land needs, nor grid integration problems, are a major hurdle to solar PV systems having a high penetration in future energy systems. Global cumulative PV capacity exceeded 1 TW in 2022, with distributed PV systems representing about 44\% of both the annual and the cumulative installed capacities in 2021 \cite{IEAPVPS_2021}. Although utility-scale PV is expected to be the major solar power source in many countries, distributed PV systems should not be overlooked as they have unique advantages that could benefit deep decarbonization. 

Distributed or rooftop solar PV, is situated within the distribution network on rooftops, parking lots, or nearby consumers, while centralized or utility PV plants are connected to transmission network and located in regions where solar potential and interconnection capacity are high. Each installation type has its own pros and cons. Utility PV plants are more cost-efficient due to economies of scale and compatibility with current power management systems. However, they may require expensive additional transmission capacity. Distributed PV offers the advantage of proximity to demand, reducing power transfer needs. However, it may introduce reverse currents and operational uncertainties for distribution grid operators \cite{wilkinson2021rooftop, van2018pv, burger_2019}. The key advantage of distributed PV is its easy integration into existing infrastructure, beneficial for constrained transmission or distribution networks with high power losses. The raise in distributed generation can balance the expected increase in distributed electricity demand from electric vehicles (EVs) and heat pumps. 

Besides technical advantages, distributed PV systems have many social benefits. Generally, they have more favourable public attitudes than wind turbines and ground-mounted PV plants. They are easily accessible, increase grid security for consumers, and empower individuals to contribute to climate change mitigation, thereby fostering additional investments by the public for the energy transition. Studies have shown that distributed PV systems can serve as a tool for raising awareness, promoting green practices, and introducing collective and distributed self-consumption models \cite{IEAPVPS_2021}. Furthermore, they have the potential to alleviate energy poverty by providing reliable, sustainable, and increasingly affordable electricity in under-resourced communities \cite{ o_2021, o_2021impact, fox_2023}. An energy system model cannot capture all of these social features, but they make it worthwhile to investigate the impact of distributed PV systems more carefully.
 
Distributed PV integration and interaction with the grid has been a subject of extensive research. Numerous studies have analyzed the effects of PV on grid voltage, frequency, and costs \cite{ gupta_2021, alboaouh_2020, karimi_2016, gandhi_2020, olowu_2018} . However, there is high uncertainty in the results of such studies due to the complexity of distribution grids \cite{ burger_2019, horowitz_2018distribution} , with some showing very little effects due to PV, others showing network upgrade costs up to 100 €/kW for high PV penetration levels \cite{lumbreras_2018, horowitz_2018, cohen_2016effects, cohen_2016effects2} , and how these PV integration costs could be lower \cite{gupta_2021} , or higher \cite{fernandez_2010} , for urban areas compared to rural ones. Uncertainties arise from the fact that studies are usually using real grid data on a small-scale and the value of distributed PV in a system is particularly location-dependent. Some studies create uncertainty by mandating a specific PV capacity, instead of optimizing it, when calculating integration costs. However, getting more robust results is becoming an urgent need as for example the question of grid costs for rising PV penetration is already steering some countries towards specifying tariffs that diminish distributed PV's competitiveness. Other questions are which concepts, such as energy communities, or distributed storage, should be supported by new schemes to pave the way for distributed PV development \cite{IEAPVPS_2021}.

Macro-energy systems models are used for large-scale system analyses extending across countries or continents. Utility PV and distributed PV systems are respectively connected to high-voltage (HV) and low-voltage (LV) levels of the grid. Many studies solely focus on modeling the system at the HV level, assuming a lossless connection from transmission to distribution grid. Low resolution models overlook congestion in transmission lines and diminish potential cost reductions resulting from local generation. Distributed PV analysis, however, necessitates distinguishing between the HV and LV level, while also considering high spatial resolution. Additionally, the scale of the analysis plays a crucial role as examining smaller areas can yield non-universal outcomes, as discussed earlier. While modeling the power grid at a continental scale with consumer-resolution is computationally infeasible, several studies have explored methods to incorporate distributed PV in macro-energy system models.

One approach is to first calculate the distribution costs or optimal technology mix for the LV or HV level of the grid, and then use the results in the other level. For instance, in their first step, Hess et al. \cite{hess_2018}  calculated distribution grid costs using a meta-analysis of previous studies, and an energy system analysis with a 491 node model. They then used the obtained costs in a one-node model representing Germany to assess grid expansion need for a 100\% renewable system by 2050. In another bottom-up approach, Baecker and Candas  \cite{candas_2022}   developed a co-optimization framework for the transmission and distribution levels of the German energy system. This framework included the development of several typical urban and rural microgrids connected to the transmission grid with an interface, being modeled in a period of 10 days to reduce the computational complexity. The results showed a positive correlation between low-cost distributed PV and the smart operation of heat pumps and EV charging stations, indicating the importance of flexibility in the low voltage grid. In a top-down approach, Müller et al. \cite{muller_2019}  first modeled the HV level of the German grid, then used the results as input for optimization of curtailment and storage units installed in LV level. They concluded that HV results limited optimization on the LV level. Clack et al. \cite{clack_2020}  employed a parametrization method to divide the utility and distribution grid by using substations as an interface where electricity flow would cross the boundary of one grid. This interface has the option to include separate costs for inflow and backflow and allowed the simultaneous optimization of utility and distribution infrastructure. This approach was employed for optimizing the energy system of the US from 2020 to 2050, and different scenarios showed cumulative cost savings of up to 18\% from distributed generation by 2050. 

Child et al. \cite{child_2019} also followed a two-steps optimization process. First, distributed technology capacities were determined with the goal of minimizing electricity cost for prosumers, then the capacity of other technologies is optimized with the goal of minimizing the cost for the whole system. A 17\% drop in annual electricity imported from the grid, and a 6\% reduction of the peak demand was achieved by solar PV prosumers. Finally, the TYNDP 2022 scenario report \cite{TYNDP}  assessed three scenarios for modeling the green transition of European energy grid from 2020 to 2050. The assumptions for the 'Distributed Energy' scenario included lower costs for solar PV systems and batteries, higher cost for wind, more decommissioning of conventional plants, and less gas demand than the other scenarios. The optimization for this scenario led to a higher share of renewables than the two other scenarios, in addition to higher electricity demand due to more electrification in the transport and heat sectors.

We extend the previous knowledge by addressing the research gap of modelling distributed and utility PV separately while examining their operational impact on the entire energy system. Hence, our primary research question is: Does distributed PV generate sufficient cost savings from self-consumption and capacity deferrals in the distribution and transmission network to counterbalance economies of scale and benefit the overall energy system? Our focus is to understand the role of distributed PV in system-wide cost optimization, adopting a social planner's perspective and utilizing a model that incorporates realistic grid mechanisms. Additionally, we address the challenge of modeling distributed PV in a macro-energy system that minimizes total system cost. To achieve this, we employ a simplified approach, which is elaborated in the Methods section.

To summarize the following study, we implement a stylized model of the distributed PV that enables to co-optimize its capacity together with other system components in a highly renewable European energy system with 181 nodes and two hourly resolution. This study represents a novel approach as it incorporates high-resolution modeling of the distribution grid and analyzes the role of distributed PV in the European Energy system. We investigate: (i) the effect of distributed solar PV on costs, components, and operation of the system; (ii) the effect of distribution grid costs and losses on the capacity and operation of distributed solar PV, and (iii) the relation between distributed solar PV and distribution grid with other system components such as transmission network and storage. We show that including distributed PV in a cost-optimal European energy system leads to a a cost reduction of 1.4\% for the power system, and 1.9-3.7\% when the complete sector-coupled system is analyzed. This is because, although distributed PV has higher costs, the local production of power reduces the need for HV to LV power transfer. Distributed PV is utilized alongside home batteries and predominantly installed in conjunction with a  higher share of EVs and heat pumps connected to the distribution grid.

\section{Methods}

We model a future European energy system with global $\mathrm{CO_2}$ emissions limited to 5\% of 1990 level, using 2-hour time resolution for a full year, and 181 nodes to represent the different regions (Figure 2). We co-optimise distributed PV generation and investment together with the entire energy system, including generation, storage, transmission, and distribution. We model the configuration shown in Figure 1 by defining two buses: HV and LV. Each HV bus is connected to HV buses through AC or DC transmission lines, and to the LV bus of the same node. The connection from the HV to the LV bus represents the lumped capacity of distribution grids in each node, a capacity which can be extended if deemed cost-effective. This way, the capacity of the distribution network is part of the optimisation. Distribution grids for each region are aggregated into one bidirectional connection based on the assumption that all the individual grids within a node, for example those corresponding to different municipalities within a country, have similar levels of PV and electric vehicles penetration. Basically, distributed technologies are assumed to be equally prevalent in close regional based on community perceptions and government policies.

\begin{figure}[!htb]
    \includegraphics[width=1\textwidth,]{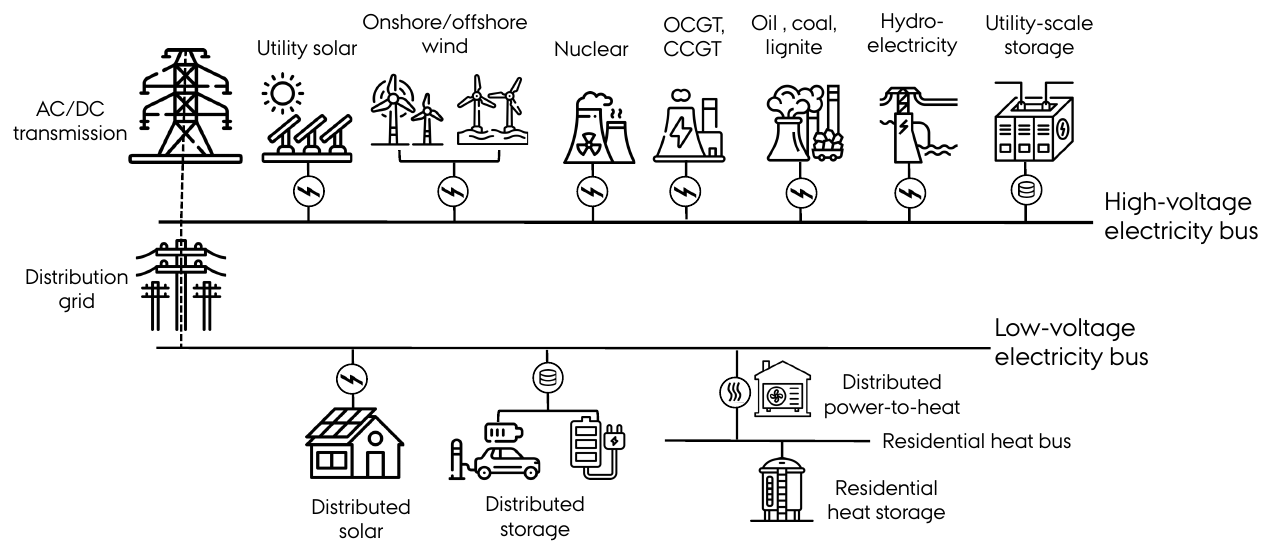}
   \caption{\textbf{} Representation of technologies and their connections to the high-voltage and low-voltage buses}
   \label{fig:pypsa}
\end{figure}

There are two main features of the distribution grid that are important: cost and power losses. These parameters depend on the distribution grid's physical properties and other factors like population density and urbanization level \cite{sadovskaia_2019}. Table 1 shows a review of distribution costs and losses reported in the literature. The cost range is 265-1350 €/kW, and the average losses are 6-8\% for European countries. Some of the differences in costs estimation are due to different assumptions regarding discount rate, equipment lifetime, rural or urban location, and how the costs of other assets such as land rights, towers, and salaries change compared to initial investment cost \cite{gupta_2021,rauschkolb_2021}. The costs in Table 1 are only 'grid extension' costs, as in how much it would cost to increase the distribution grid capacity in 1 kW. There are also studies that estimate costs for 'grid reinforcement', as in upgrading distribution lines or transformers, which could be useful to gauge how much building the network from scratch would cost. Some of the results of these studies are summarized in SI Table S1. 
 
\begin{table}[]
\caption{ Overview of costs and losses assumed for electricity distribution
grid in previous studies}
\scriptsize
\begin{tabular}{m{3cm}m{1.5cm}m{1.5cm}m{2cm}m{1.3cm}m{1.3cm}}
\hlineB{4}
\multicolumn{6}{l}{\textbf{Distribution grid costs}} \\ \hlineB{4}
References & Capital cost (€/kW) &  Annual capital cost (€/kW/year) & O\&M (€/kW/year) & Lifetime (years) & Discount rate \\
\hline
PyPSA-Eur-Sec \cite{hess_2018} & 500 & 47.5 & 2\% of Capital & 40 & 7\% \\
Fares \& King (2017) \cite{fares_2017} Clack et al. \cite{clack_2020}  & - & 53 & - & - & - \\
Hess et al. \cite{hess_2018} (2018) & 375 - 500 & 65.5 & - & - & - \\
Zhang et al. (2021) \cite{zhang_2021} & 1351 \$ & 82  & - & 40 & 4.40\% \\
Rauschkolb et al. (2021) \cite{rauschkolb_2021} & - & 75 & - & 30 & 8\% \\
Danish Energy Agency  & 265-560 & 15-32 & 0.5\%  of Capital & 40 & 4\% \\ \hlineB{4}
\multicolumn{6}{l}{\textbf{Distribution grid losses}} \\ \hlineB{4}
References & Loss &  &  &  &  \\
\hline
Sadovskaia et al. (2019) \cite{sadovskaia_2019}, World bank Data (2014) \cite{worldbsnk_2014} & \multicolumn{5}{l}{2\%-22\% : average 8\%}   \\
Council of European Energy Regulators Report on Power Losses (2017) \cite{ceer_2017} & \multicolumn{5}{l}{1\%-13.5\%: average 6\%    (for European countries)} \\ \bottomrule
\end{tabular}
\end{table}

The optimization is performed using the open model PyPSA-Eur-Sec and assuming different combinations for the distribution costs and power losses. A detailed description of the model is provided in Victoria et al.\cite{victoria_2022}. Assumptions regarding grid costs and grid losses are based on the approximate minimum and maximum values from Table 1 to best demonstrate the effects of distribution grid features (refer to SI, figure S4). It should be noted that since transmission network has no losses in the reference scenario, we assume all the power losses happen in the distribution grid.

A summary of each scenario’s parameters is shown in Table 2. Since our main goal is to assess how distributed solar affects the system, every scenario is simulated once without any distributed PV and storage, and once with these technologies included. Our model assumes greenfield optimization for most of the technologies and costs for 2030 (refer to SI, table S2). One exception is that the wind and solar PV installed capacities in 2020 and the capacities of the existing transmission lines are imposed as lower limits. Another exception is that reservoir and run-of-river hydropower plants, as well as pump hydro storage (PHS), are exogenously fixed at 2020 capacities since their potential expansion is limited in Europe. We selected greenfield optimization under the assumption that most of the renewable capacity will be newly built in a distant future year. Our focus is on understanding the impact of distributed generation, so we select 2030 as to account for expected renewable cost reductions conservatively, not because we intend to specifically represent that year. The model includes all EU27 countries except for Malta and Cyprus, as well as Albania, Great Britain, Montenegro, Norway, Serbia, and Switzerland. Transmission expansion is allowed at only 10\% added capacity compared to today for the main scenarios, to avoid social acceptance issues.

\begin{table}[]
\caption{ Summary of scenario assumptions including distribution grid costs and distribution grid power losses. Discount rate is assumed to be 7\% for utility-scale solar PV and 4\% for distributed solar PV. Operation and maintenance is 1.95\% of investment cost for utility solar and 1.42\% for distributed solar. Lifetime is 40 years for both.}
\scriptsize
\resizebox{\columnwidth}{!}{%
\begin{tabular}{m{1.2cm}>{\raggedright\arraybackslash}m{2.5cm}m{1.6cm}m{1.5cm}>{\raggedright\arraybackslash}p{2.2cm}}
\hlineB{2}
\textbf{Scenario} & \textbf{Sectors} & \textbf{Dist. grid cost} (€/kW) & \textbf{Dist. grid power loss} & \textbf{Modes} \\
\hlineB{2}
\textbf{A} & Electricity & 500 & 0\% & \multirow{2}{3cm}{With/Without distributed solar and home batteries} \\
\cmidrule(r){1-4}
\textbf{B} & Electricity & 1000 & 10\% &  \\
& & & & \\
\hline
\textbf{C} & All (electricity+ heating+land transport+ industry+  shipping+aviation) & 1000 & 10\% & \multirow{2}{3cm} {With/Without distributed solar and home batteries \newline (With EVs and heat pumps)} \\
\cmidrule(r){1-4}
\textbf{D} & All (with high dist. solar potential) & 1000 & 10\% &  \\
\hline
\end{tabular}%
}
\end{table}

Power flows for AC and DC transmission lines connecting the nodes are modeled using a linearised optimal power flow (LOPF) formulation that captures some of the physical characteristics of the power transmission network by including Kirchhoff’s current and voltage law constraints (as described in SI). AC flows are linearised using DC linearisation, assuming that voltage angles differences across branches are small, and that branch resistances are negligible compared to reactance, which means that power losses at the transmission level are neglected \cite{brown_2017,neumann2022assessments,pypsa_2016}. The energy balance constraint ensures that at every time-step, node and sector the inelastic demand is met. 

The connections between HV and LV buses representing distribution grids in every node are modeled with controllable power flow using a transport model, and these connections could include a constant efficiency to represent power losses. Using the transport model, we ignore congestion in the distribution grid. This assumption arguably favours utility technologies as they have higher energy generation and cause congestion more frequently. In every node, utility-scale technologies and transmission are connected to the HV bus, while demand and distributed generation technologies are connected to the LV bus (see Figure 1). The grid connection cost for utility technologies is assume to be zero, another assumption which would enhance distributed generation profitability if implemented.

The first two scenarios, A and B, represent the electricity sector. In this case the electricity demand also includes the heating demand that today is provided by electric appliances. The power sector model includes onshore and offshore wind, utility-scale and distributed solar, run of river, open-cycle and closed-cycle gas turbines, hydrogen fuel cell, nuclear, coal, and oil power plants as electricity generation technologies. Electricity storage technologies included in the model are utility-scale batteries, home batteries, pumped hydro storage, and hydrogen storage. The sector-coupled scenario adds separate demand and technologies for heating, land transport, aviation, shipping, and industry sector including feedstock to the model. The added technologies for power and heat generation are solar thermal collectors, heat pumps, resistive heaters, gas/biomass boilers, and gas/biomass combined heat and power plants (CHP). Added storage technologies include electric vehicle batteries and water tanks. A detailed description on other technologies such as carbon capture and how the different sectors are modeled can be found in Victoria et al. \cite{victoria_2022}. Technologies connected to the LV bus are distributed solar, home batteries, electric vehicle batteries, and heat pumps.

Utility PV cost is assumed to be 347.6 €/kW, and distributed PV cost is assumed to be 636.7 €/kW \cite{dea_2020}. To model the maximum solar utility potential, 9\% of the available land suited for ground PV installation is considered. This available land is based on the Corine Land Cover database and discounts non-suitable land categories such as natural reserves, cities, etc. No cost is considered for connecting utility-scale PV plants to the grid. For distributed PV, a potential of 1 kW per capita is assumed based on 10 m\textsuperscript{2} of rooftop available per citizen, 50\% of which is considered suitable to install PV, with 20\% efficient PV modules. This leads to a total of 504 GW potential for distributed solar in Europe that could produce almost 550 TWh/a, a number close to results from other detailed analysis \cite{huld_2018,bodis_2019}. Total utility-solar potential and maximum generation is 10.5 GW and 11.7 PWh, respectively. Assumptions for both utility and distributed PV potential are conservative compared to other research conducted in this area \cite{trondle_2019}. Figure 2.a and 2.b show the regional electricity demand and the potential distributed PV capacity, respectively; while Figure 2.c shows the correlation between the two parameters. As distributed PV capacity is calculated based on population, most areas in Figure 2.c are in the diagonal of the color plot. Many areas have high distributed PV potential, but it might still only cover a small percentage of demand, especially if the area has high industrial electricity demand. Figure 2.d shows the percentage of electricity demand that could be provided by distributed solar in each region if the estimated potential are fully utilized. 

\begin{figure}[!htb]
    \includegraphics[width=1\textwidth,]{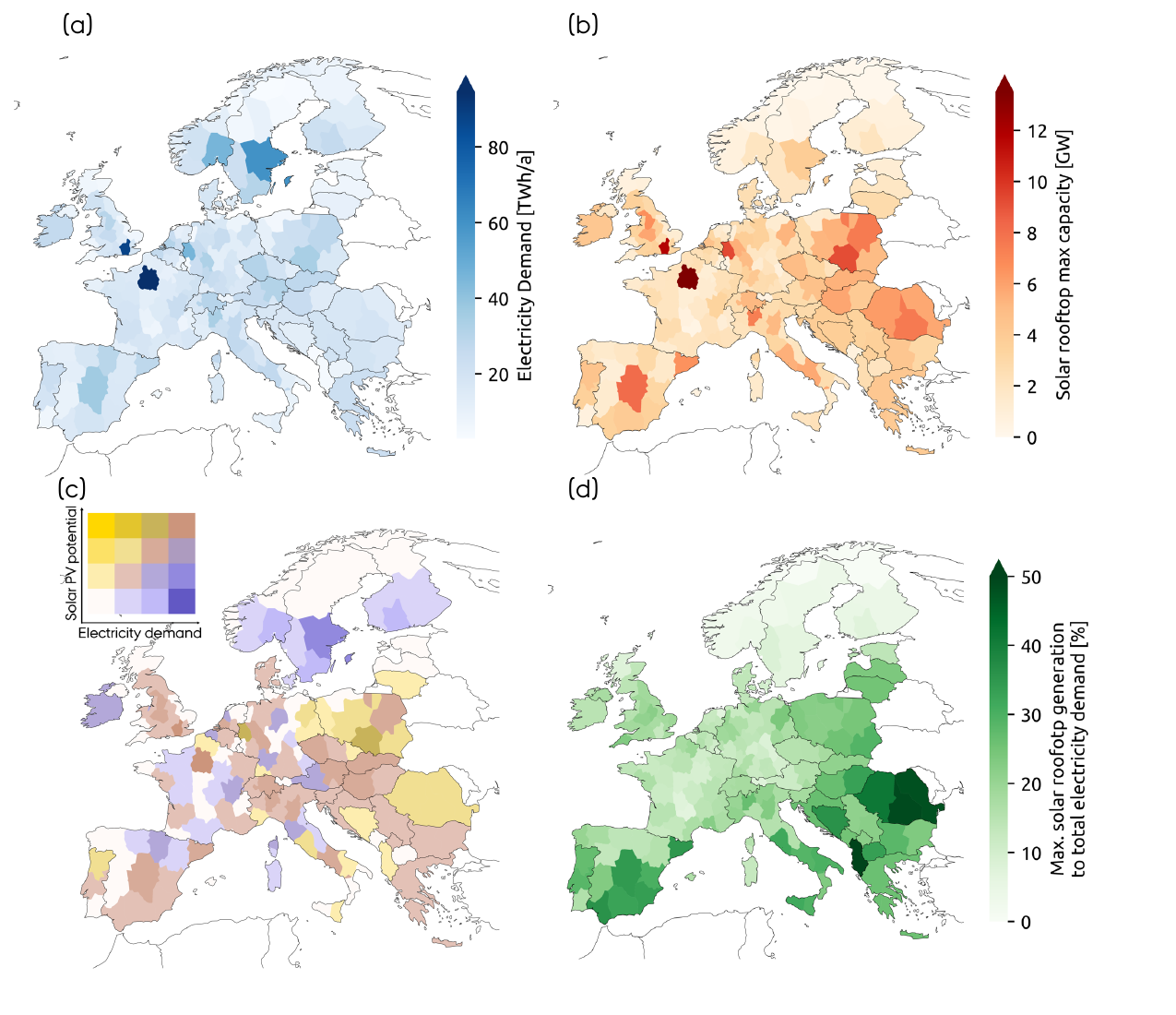}
    \caption{\textbf{Map for the 181 regions considered in the model displaying: a) Electricity demand, b) Distributed solar PV potential, c) Correlation of electricity demand and distributed solar potential, and d) Ratio of maximum distributed solar generation (assuming all capacity is installed) to electricity demand (\%)} .}
   \label{fig:demand/capacity}
\end{figure}

\section{Results and discussion}

\subsection{Trends in system costs and capacity}

Total system costs for the three scenarios, with and without distributed generation, are shown in Figure 3. For all scenarios, distributed generation reduces the total system cost. For scenario A, with no power distribution losses and investment costs for the distribution grid of 500 €/kW, the overall system configuration and costs remains roughly the same whether distributed generation is included in the system or not. Scenario B assumes 10\% distribution power losses and cost of 1000 €/kW for the distribution grid, showing overall system costs higher than scenario A due to additional generation required to compensate the distribution grid's losses. For scenario B, total system costs decrease by 1.44\%, corresponding to 4.1 billion euros per annum, when optimization includes distributed PV generation and home batteries. In scenario A, the system prioritizes utility solar due to the minimal costs associated with transferring energy to the demand bus. Therefore, the absence of distributed PV in scenario A does not affect costs noticeably. However, when faced with a 10\% power loss and higher costs for energy distribution, the system makes the decision to utilize technologies directly connected to the LV bus. Therefore, the low cost and no power loss assumptions for power transfer, used by default in many models, hinder the selection of distributed generation in the optimal solution.

For the sector-coupled scenario C, where 10\% power losses and 1000 €/kW for the distribution grid are assumed, the cost reduction reaches 14.1 billion euros per annum, equal to a 1.9\% reduction. The distributed PV potential is fully utilized in scenario C, so an additional scenario D with 6-times distributed PV potential, equal to 3 TW, is also modeled. The total cost savings for scenario D from distributed PV reach 3.7\%, by installing 2.1 TW of distributed PV. This could represent a future scenario in which distributed PV is particularly favored by policy and the dual use of infrastructure extends not only to residential buildings, but also to industrial, commercial, and public buildings, parking lots, ground mounted systems in urban environment, etc. \cite{trondle_2019, France_senate2022}. The total costs of technologies using gas for heat production is increased in scenarios C and D when distributed solar is included. This is mainly caused by an increase in gas boilers which will be further discussed later in the paper. It should be noted that the costs for electric vehicles in scenarios C and D are not included.

\begin{figure}
   \includegraphics[width=0.95\textwidth,]{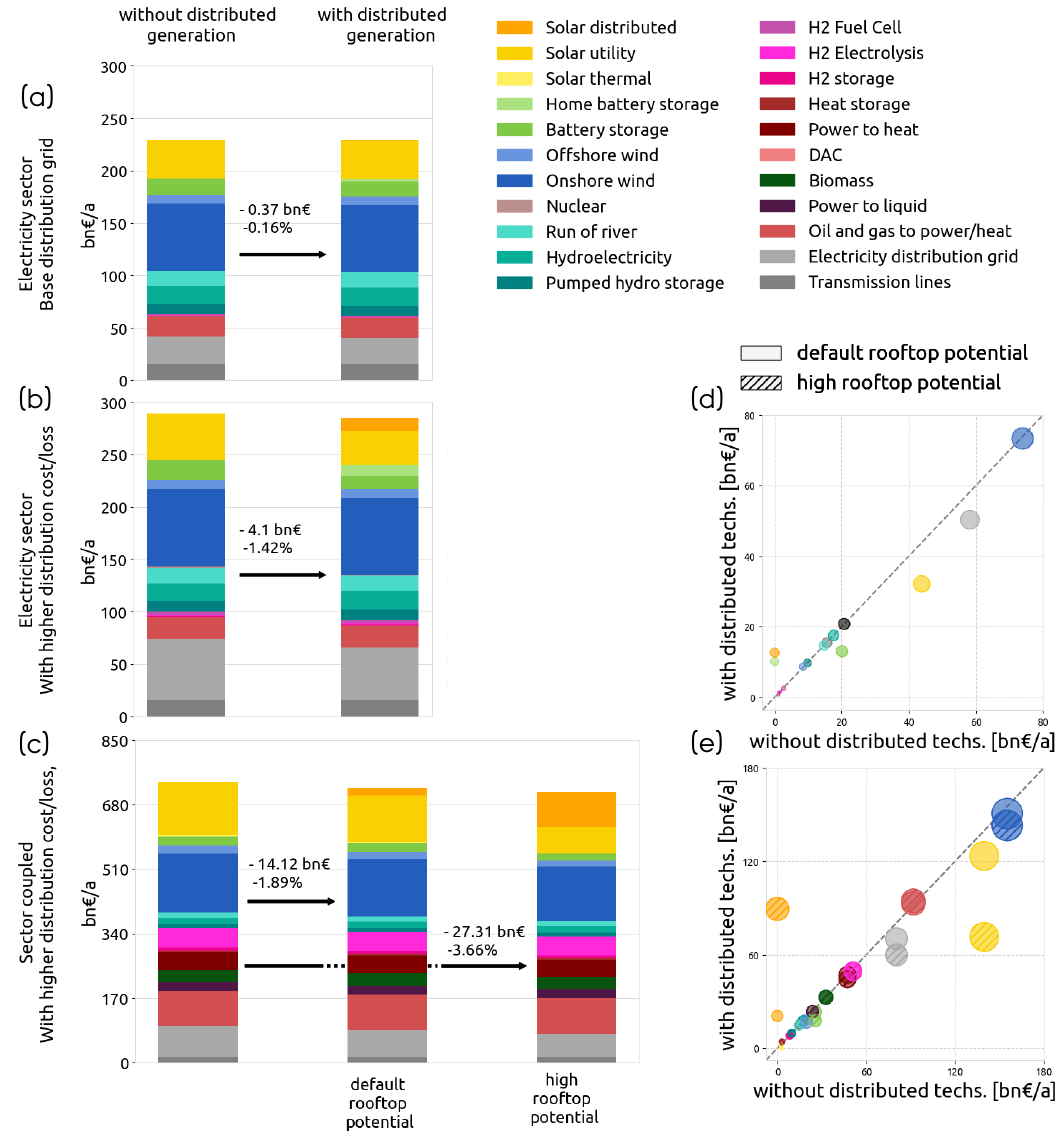}
   \caption{\textbf{Annual total system costs composition for each scenario with and without the inclusion of distributed solar and home batteries in the system shown in (a), (b), and (c). The cost reduction with distributed generation, and the reduction percentage relative to the system without distributed generation, is shown above arrow marks for each scenario. Cost change in scenarios B, C, and D  are also represented in plots (d) and (e), respectively, to show a more clear comparison of changes in total investment costs for each technology.}}
   \label{fig:costs}
\end{figure}

Higher capacities of distributed solar and home batteries in the system means lower capacities of utility solar, utility batteries, and distribution grid, as shown in Figure 4. Other major technologies like wind energy experience no significant difference without distributed generation, and their changes in capacity are below 5\%. The effects of distributed generation are much more visible in scenarios B and C. For scenario A, there is respectively a 0.7\%, 10.5\%, and 5.1\% reduction in solar utility, utility batteries, and distribution grid capacity when system is optimized with distributed generation. The small solar utility capacity reduction is almost equal to added capacity of distributed solar, which is installed primarily in southern Europe near nodes where solar utility is already installed at maximum capacity. The utility battery power capacity reduction is replaced by 1.5 times the capacity added as home batteries. This merely indicates that using home batteries to reduce the demand peak in distribution grid is cost efficient.

The three capacities experience a significant reduction for scenario B, being equal to 26.4\%, 33.8\%, and 13.5\% for solar utility, utility batteries, and distribution grid, respectively. Again, reduced capacity of utility solar and utility batteries is almost equal to the added capacity of distributed solar and home batteries. For the sector-coupled scenario, the changes are more significant if we assume a higher potential distributed PV, being equal to 48.8\%, 40.3\%, and 25.5\% for solar utility, utility batteries, and distribution grid, respectively. 

While distributed solar capacity is only 1.6\% of the maximum potential for scenario A, it shows a staggering increase to 60.9\% for the scenario B, in which 307 GW of distributed PV are installed, and 99.9\% for scenario C, in which 504 GW of distributed PV is installed. Increasing the potential in scenario D leads to 2170 GW of distributed PV installation, meaning more than half of the solar generation is installed in distributed form. This scenario might seem extreme in terms of the very high installed capacity of distributed solar, but it shows that moving towards a fully distributed system would still be cost efficient. A sensitivity analysis for this scenario is included in SI (Figure S29) assuming lower costs and power losses for the distribution grid, indicating the share of distributed PV still remains significant.

Distributed generation leads to lower capacities of the distribution grid if distributed storage is available. For scenarios A and B, distribution grid capacity is reduced by 5\% and 13.5\%, respectively. The greater reduction percentage in scenario B is attributed to its more expensive grid, which amplifies the need for cost-saving measures. Power capacity of home batteries for scenarios A and B is 5.1\% and 14.1\% of the low-voltage (LV) peak demand, respectively. The average discharge time, which is the energy to power ratio for batteries, is about 6 hours for all scenarios. As previously discussed in Victora et al.\cite{victoria_2019role} and Brown et al.\cite{brown_2018}, the presence of electric vehicle batteries, heat pumps, and better flexibility when other sectors are included in Scenarios C and D means that static home batteries are no longer needed.

\begin{figure}
   \includegraphics[width=1\textwidth,]{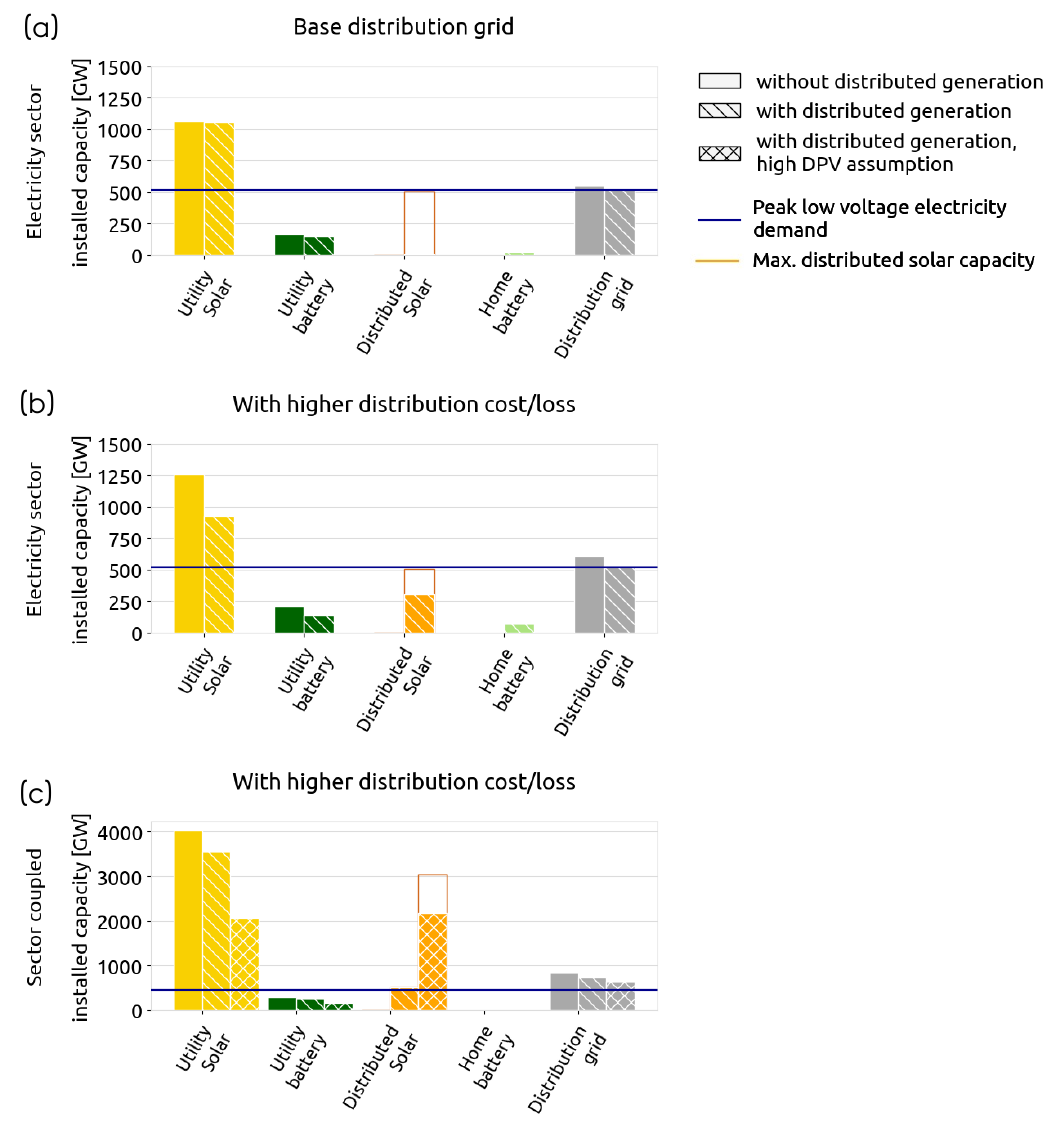}
   \caption{\textbf{Change in installed capacities of four technologies for a) scenario A (500 €/kW, 0\% power loss), b) scenario B (1000 €/kW, 10\% power loss), and c) scenario C and D (sector-coupled, 1000 €/kW, 10\% power loss). Notice that the scale in scenario C/D is different than A and B. 
   The horizontal line and the orange rectangle on each graph represent the considered total distributed solar PV (DPV) potential and peak low-voltage (LV) electricity demand in the model. LV demand includes both residential and industry electricity demand.}}
   \label{fig:capacities}
\end{figure}

\subsection{Regional and temporal patterns for distributed generation}

Having seen the overall benefits of distributed generation for the system, we can now look at where distributed generation is more prevalent and what effects it induces. Figure 5.a and 5.b show the share of solar utility and distributed solar generation on the total annual energy generation in each European country, and each node, for scenario B. We can recognize that Southern countries have more solar, both in distributed and utility form, compared to the Northern regions. The distributed solar capacity is however more sensitive to the available solar resource since it is more expensive than utility solar and needs higher radiation levels to become cost-efficient. 

Energy generation mix in places with a high share of distributed generation shows a visible reduction in distribution grid usage during summer. Figure 5.c shows the energy mix and demand of Spain in the winter and summer weeks for the scenario B. Only at times when all electricity demand is covered by utility generation, total demand (black line) equals the LV demand (red line). The summer week shows a stable pattern of utility generation from different technologies. As already observed in similar studies\cite{candas_2022,clack_2020}, the LV demand becomes flattened as distributed solar and home batteries are added to the system. The flat shape is due to a shift towards more distributed generation which is more cost-effective, and the system attempting to minimize distribution grid capacity costs while maximizing the usage of that installed capacity. During the winter week, most of the demand is met through technologies connected to the HV bus. Although the overall contribution of distributed generation is lowered in winter, home batteries play a big role in reducing distribution load as they store energy throughout the day and release it at the end of the day when demand is ramping up and solar production is decreasing.

\begin{figure}
   \includegraphics[width=1\textwidth,]{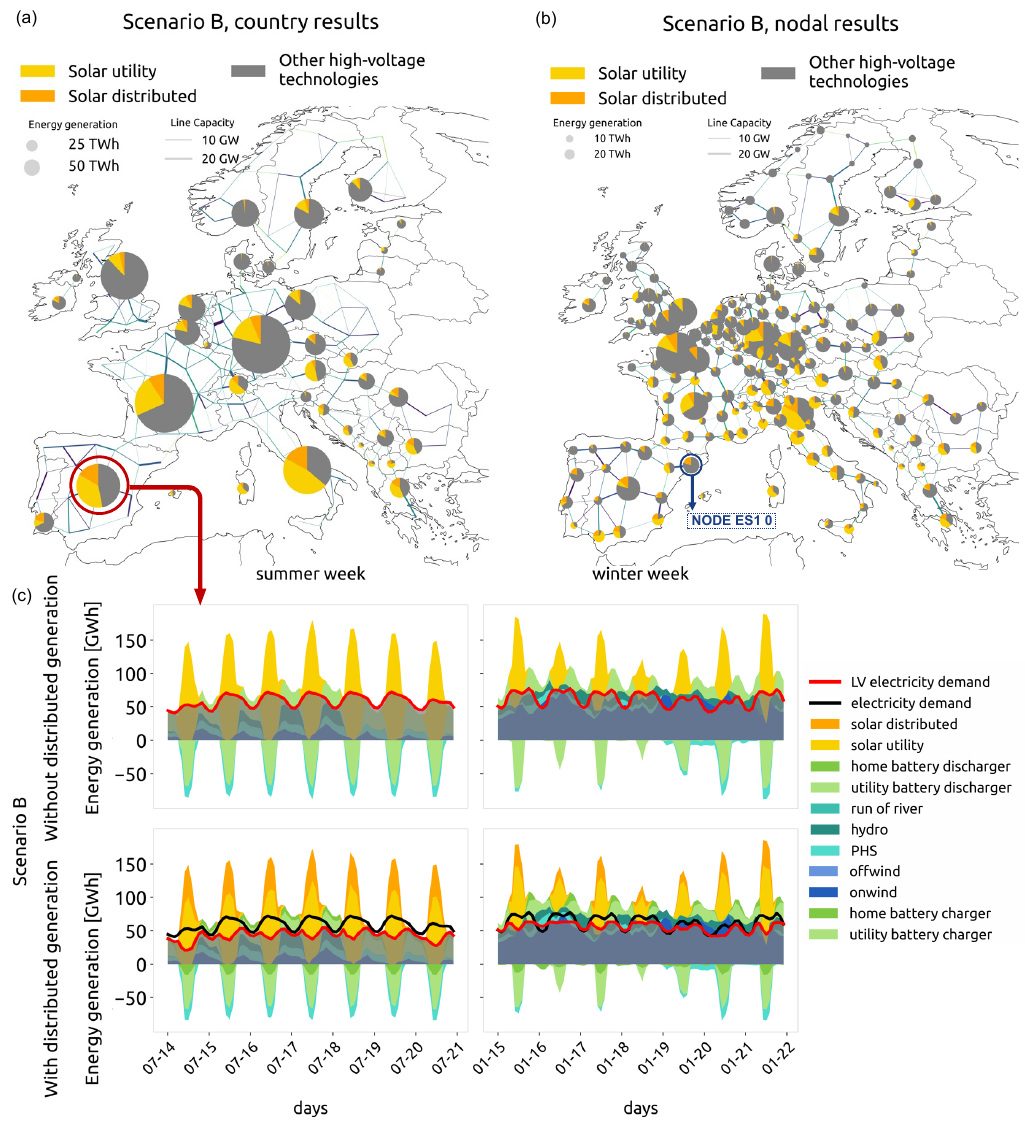}
   \caption{\textbf{ Solar utility, distributed solar, and other technologies (wind, nuclear, etc.) share of annual electricity demand for scenario B in a) each country and b) each node. c)  energy generation mix of Spain for a week in Summer without (top) and with (bottom) distributed solar and batteries for scenario B. }}
   \label{fig:maps}
\end{figure}

The decrease in HV to LV energy transfer is the main factor differentiating systems with and without distributed technologies. In the following we analyse this effect. There is both an overall peak reduction in LV demand during the year, and a daily reduction happening during daytime. Figure 6 shows the energy transfer through the HV-LV connection representing distribution grid for an example region in Spain. In this region, marked in Figure 5.b, distributed solar share of the annual generation is 18.9\%. The top four figures compare the yearly and average daily energy transmission from HV to LV and vice-versa. Without distributed generation, the entire electricity demand is being met by generation at the HV buses, so the demand coincides perfectly with the HV to LV energy flow. When distributed generation is included, the HV to LV energy transmission is greatly reduced, especially during summer, indicating a good amount of self-consumption. In some rare cases during the summer there is even some energy being transferred from LV buses to HV buses, shown in blue, which is energy from distributed solar that could be transmitted to other nodes or stored in utility storage. The daily pattern of energy flowing through the distribution grid also shows how the average daytime distribution peak load is reduced nearly 60\% in the middle of the day, showing the characteristic 'duck' curve \cite{denholm2008production,  caiso2016duck}. There is even reduction at night-time, which is most likely due to self-consumption from home batteries.

For the example region in Spain, there is at least a 20\% reduction in distribution grid load during 50\% of the year. Figure 6.c compares the duration curve of the energy flow through distribution grid from in the example region of Spain with and without distributed generation. The overall curve has shifted downwards, which is the peak reduction that was mentioned, and the shape takes a sharper turn downwards at the right side, which shows the more drastic peak reductions that are happening during the day in summer months. The curve also has negative values for scenario B with distributed generation, which is the energy being transferred from LV to HV buses. Additional duration curves for different node groups are included in the SI (Figure S19).
\begin{figure}
   \includegraphics[width=1\textwidth,]{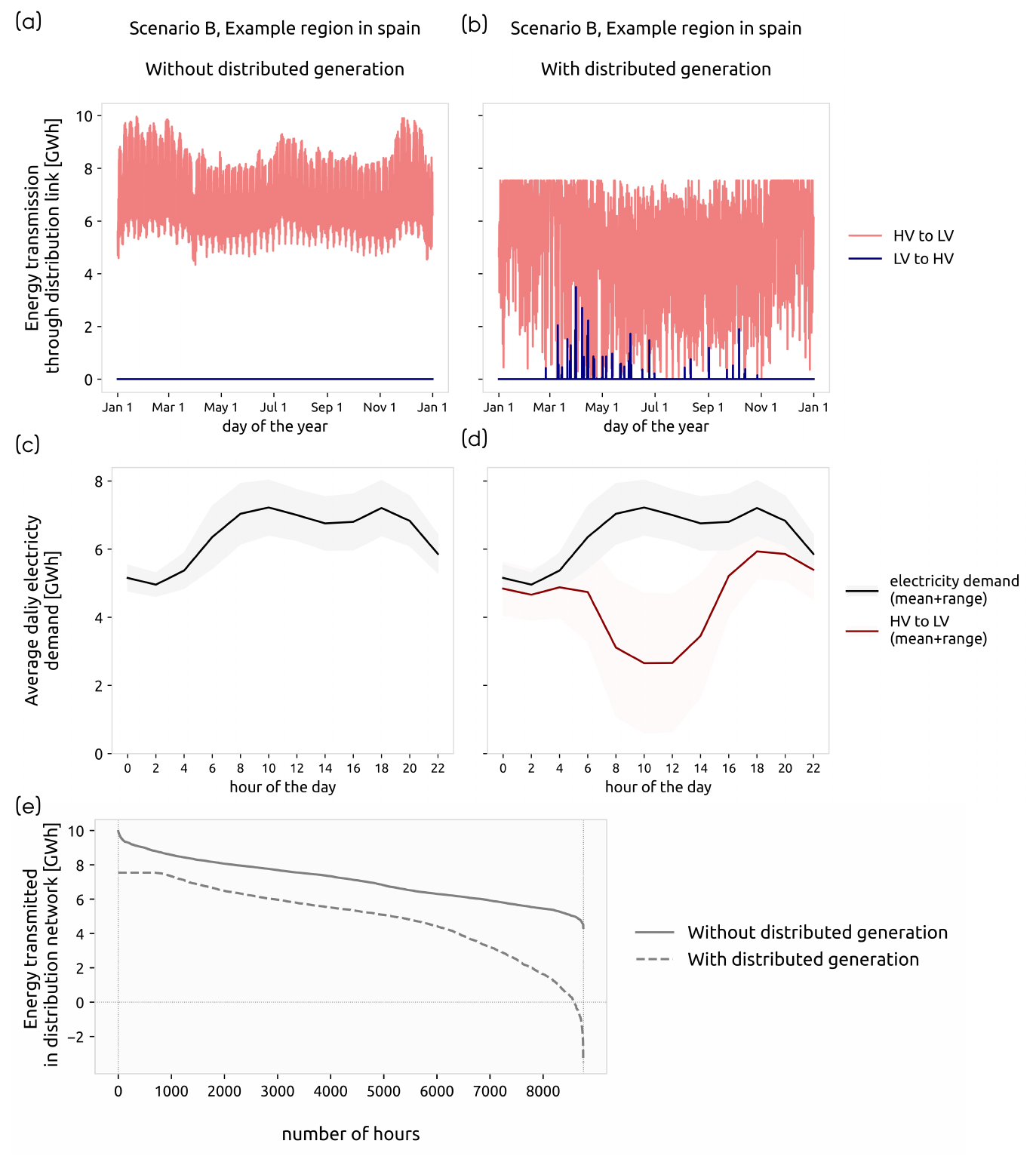}
   \caption{\textbf{Energy transfer through the HV-LV connection representing the distribution grid for example region in Spain a) without and b) with distributed generation. c) and d) Daily pattern (averaged over the year) of electricity demand and HV to LV energy transfer which is equal to LV electricity demand. The colored area around each line represents the variability of the curve during the year. e) Duration curve}}
   \label{fig:duration}
\end{figure}

As previously mentioned, one of the advantages of distributed generation is the possibility to increase energy self-sufficiency via self-consumption of solar energy. Energy self-sufficiency is defined as the ability of a country or region to fulfil its own energy needs. Metropolitans nodes could specially benefit from this, as they have large populations with highly concentrated demand, but lack suitable land for installation of utility scale wind and solar plants, leading to a complete reliance on imports from neighbouring nodes. To evaluate how distributed solar could increase the self-sufficiency, Figure 7 shows the nodal electricity annual balance for scenario B. The green areas are net exporters generating an excess of energy annually while purple areas represent energy deficiency on an annual basis. The right figure, when distributed generation is available, shows a consistent conversion of colors towards white for southern Europe due to higher distributed solar potential. There is also a visible shift observable for the Paris and Madrid nodes as they reduce electricity imports, but such an observation is not discernible for the London node.

Together with the presence of distributed technologies, the other factor highly impacting on regions self-sufficiency is the expansion of transmission capacity among the nodes. The Gini diagrams in the subplot show how electricity balances change when the system has higher transmission expansion allowance. In the reference case, transmission capacity through HV lines can only be expanded by 10\% of current volume. Increasing transmission allowance for scenario B leads to an overall cheaper system and lower capacity installation for all technologies except for wind (refer to SI, figure S16). Local generation and the expansion of transmission cause opposite effects. Strong local generation reduces the need for transmission grid expansion, while a strong expansion of the transmission grid allows for greater centralised power generation at locations with good renewable resource and reduces distributed generation. The effect of moving away from local generation in presence of a bigger transmission grid is a higher overall unbalance rate for the system. A look at the Gini diagrams clearly shows a further deviation of the electricity balance from the diagonal line for expansion of transmission up to 50\% of the base network, and for ‘no-limit’ expansion that allows the system to optimise transmission capacity without limit. However, the system cost savings from distributed generation and storage remain constant, varying by less than 0.05\%, for 50\% and no-limit transmission expansion allowance scenarios. This means the contribution of distributed solar for the power sector is robust for different transmission expansion limits. 

\begin{figure}
   \includegraphics[width=1\textwidth,]{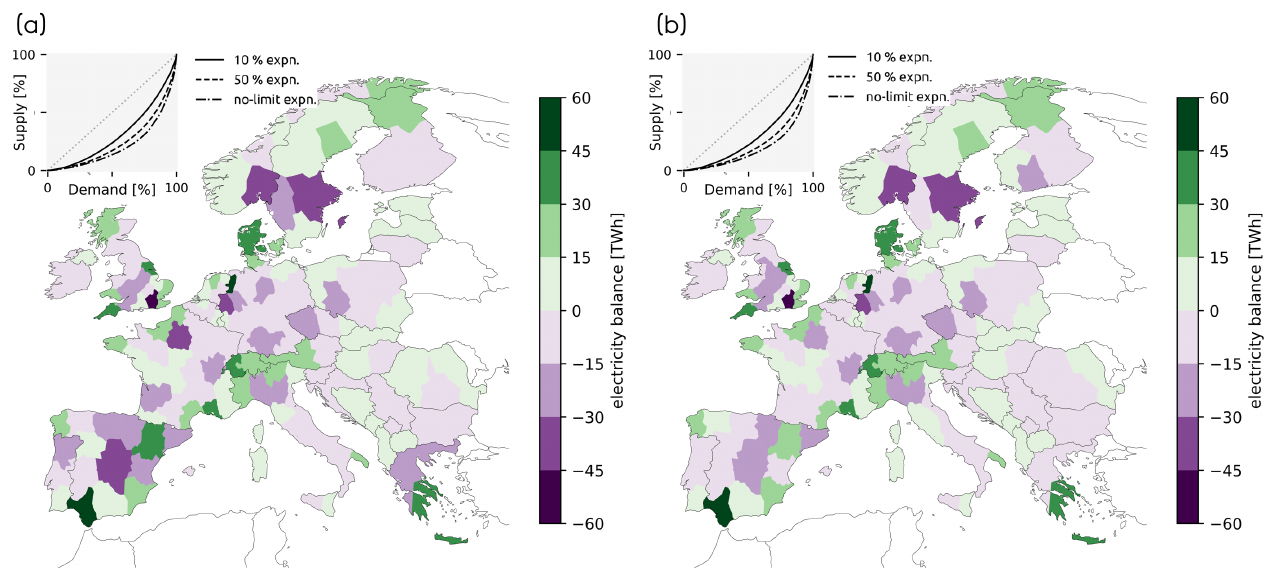}
   \caption{\textbf{Regional annual electricity balance for scenario B: a) Without distributed generation and storage, b) With distributed generation and storage. The change in overall system electricity balance when transmission network capacity is allowed to expand 10\% (default assumption), 50\%, and without limit is shown in the Gini diagrams of each figure.}}
   \label{fig:balance_maps}
\end{figure}

\subsection{Role of distributed storage}

Since solar generation is inherently intermittent, a question can be raised as to how much distributed generation and distributed storage rely on each other. Figure 8.a shows the total energy generation from distributed solar and discharging of home batteries for scenario B during the highest-demand weeks of summer and winter. It is noticeable that home batteries play a significant role in winter, as for example they discharge energy equivalent to 77\% of the distributed solar energy generation in this particular week with very high demand.

Figure 8.b shows the average daily charging and discharging for home batteries. The pattern in summer is as expected, with all the charging happening during the day from solar production. In winter, there are hours in the night when batteries are charging. This shows that due to the higher cost and losses of the distribution grid, energy is transferred to the LV buses at night and stored in home batteries so that the discharge could help reduce the LV demand peak during the day. This charging pattern is beneficial for the system as a whole. The behavior shown here can not only be provided by home batteries but also by larger batteries connected to LV buses that are managed by a distribution system operator (DSO) who aims at lowering electricity imports from HV and reduce power distribution costs.

Figure 8.c shows results for an additional simulation without distributed storage. In this case, distributed PV almost fully disappears from the system. This result emphasizes the need to develop distributed storage (either static or provided by EV batteries) to attain a high deployment of distributed solar. 

\begin{figure}
   \includegraphics[width=0.8\textwidth,center]{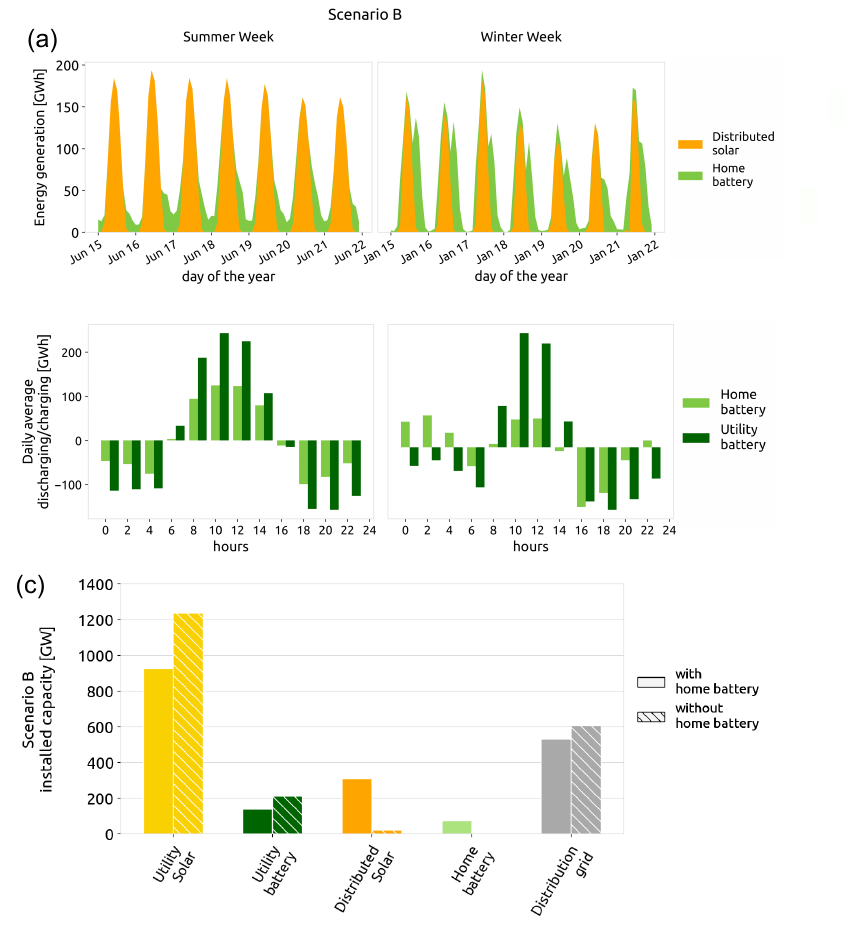}
   \caption{\textbf{a) Energy generation from distributed solar and discharging of home batteries during the highest demands weeks in summer and winter for scenario B. b) Charging and discharging of utility and home batteries during the highest demands weeks in summer and winter for scenario B. c) Installed capacity of solar utility, utility batteries, distributed solar, home batteries, and the distribution grid for scenario B with and without the presence of home battery in the system}}
   \label{fig:battery}
\end{figure}

\subsection{Impact on heating technologies mix}

Widespread electrification across all sectors is required for decarbonisation of energy systems, and there is an expected growth in distributed technologies such as heat pumps and electric vehicles options \cite{IEA_globalEV}, which can benefit from distributed PV generation since they are connected to the LV bus and in turn provide balancing for that generation. A question can raised on whether and if distributed solar impacts the optimal capacities and operation of heating technologies. 

Figure 9.a shows how the total energy generation of heating technologies changes for scenario C and scenario D when distributed solar is present. Heat pumps are the main technology providing heat both in district heating systems and individual homes, but an additional technology is required to cover the peak demand (this function is typically provided by CHP units, gas boilers, resistive heaters or thermal energy storage previously filled with solar thermal).  In previous analyses, solar thermal was not found to be competitive \cite{victoria_2022, zeyen2021mitigating}. This holds true here, assuming zero power loss and a base cost for the distribution grid (refer to SI, figure S28). 
In scenario C, about 100 TWh of solar thermal is used in district heating systems. This is because the high costs and power losses associated with the distribution grid reduce the competitiveness of CHP units connected to the HV bus. As the usage of CHP units decreases, gas boilers usage increases to compensate for the demand. However, their expansion is limited by the $\mathrm{CO_2}$ constraint, thus making solar thermal competitive. 
In scenario D, the large solar rooftop capacity provides cheap electricity at the LV bus during the day, which during peak heat demand events can be transformed into heat even using low-efficiency resistive heaters connected to district heating systems, preventing the selection of solar thermal in the optimal solution. The observed reduction in installed capacity of heat pumps in scenario D, despite their role in balancing distributed PV generation, may seem contradictory. However, Figure 9.c, illustrating the yearly energy generation from heat pumps, clearly demonstrates that heat pump generation experiences greater fluctuations during summer months when distributed solar is incorporated into the technology mix.
Both scenarios C and D have lower heat price for consumers when distributed solar is included in the system, as shown in Figure 9.c for scenario D. 

\begin{figure}
   \includegraphics[width=1\textwidth,]{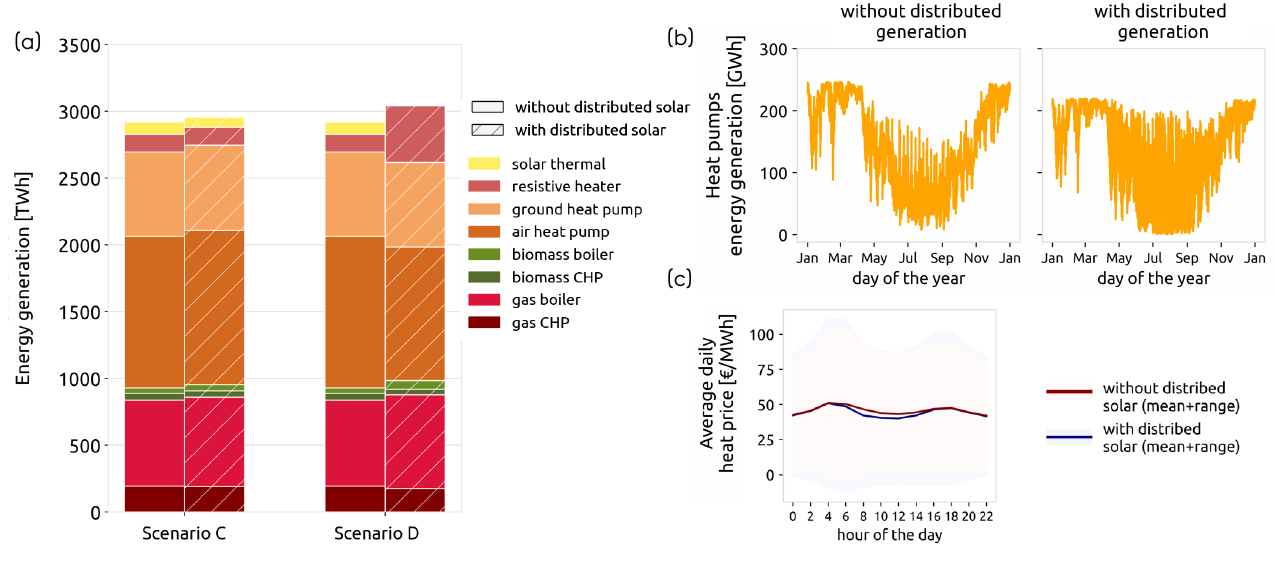}
   \caption{\textbf{a) Changes in capacity mix of heating technologies with and without distributed solar for scenarios C and D with default and high distributed solar potential b) Average daily heat price with and without distributed solar for scenario D c) Changes in yearly energy generation pattern for heat pumps with and without distributed solar for scenario D }}
   \label{fig:heat}
\end{figure}

\subsection{Comparison with relevant literature}

Although some of the quantitative results from this study strongly depend on modeling assumptions (refer to SI, figure S4), it is possible to do a general comparison of these results with other similar studies. Child et. al \cite{child_2019} results show close to 1650 GW of installed solar capacity for the power sector scenario, which is fully decarbonized by 2050. This is comparable to 1250 GW solar from scenario B with 95\% decarbonization goal. However, the share of distributed solar of total solar capacity in our study, equal to 25\%, is much lower than the 72\% obtained by Child et al. This is due to the fact that prosumer's profit is prioritized over total system costs in the aforementioned study. The TYNDP report \cite{TYNDP} includes a Distributed Energy (DE) scenario which has cost assumptions that provide favorable conditions for solar PV, with the goal of 100\% decarbonization for all sectors by 2050. Results show a total distributed PV installed capacity of 561 GW in 2050 for Europe. This is close to the 504 GW of distributed PV in our scenario C, but lower than the 2170 GW obtained for scenario D. The share of solar is 28\% of the total energy generation for the DE scenario, also close to the 32\% obtained here. It is worth noting that the REPowerEU Plan  \cite{repowerEU}, published by EU commission in 2022, includes a target PV capacity of 750 GWDC (corresponding to 600 GWAC) for Europe by 2030. The study by Clack et. al \cite{clack_2020} is conducted for the U.S. with 5 year time-steps from 2020 to 2050 to reach a 95\% decarbonized power sector in the clean energy (CE) scenarios. Comparison of results shows the share of installed capacity of distributed solar to total solar capacity, 24\%, is very close to the 25\% obtained here. However, the cumulative savings when distributed generation is co-optimized with other technologies in scenario CE-DER, equal to 18\%, are much higher compared to the 1.44\% obtained here for scenario B. The savings for the CE-DER scenario happen mostly after 2030, so the overnight optimization method used here might be underestimating the cost savings. Another difference is the distribution grid is modeled as two unidirectional connections by Clack et al., instead of one bidirectional connection used here, which was decided based on a sensitivity analysis (refer to SI, figure S2).

\vspace{\baselineskip}
\

\section{Conclusions}

In this study, we model a highly renewable European energy system represented by 181 interconnected nodes in order to analyze how distributed solar PV affects the operation and total costs of the system. The modeling is done for a full year with 2-hourly time steps to capture both the daily and seasonal changes in demand and production. 

The results show that the presence of distributed PV and distributed storage reduces total system cost. Assuming 1000 EUR/kW and 10\% power losses in distribution grids, total system cost reduces by 1.4\% when only the power sector is included and between 1.9 and 3.7\% for the sector-coupled scenario. Local energy production by distributed PV at low-voltage reduces the need to extend power distribution infrastructure to transfer energy from utility technologies at high-voltage levels, and increases energy self-sufficiency for many regions, especially in southern Europe. The entire assumed distributed PV potential, equal to 504 GW, is installed for the sector-coupled scenario. If we assume a higher potential based on installing distributed PV also on industrial, commercial, and public buildings, parking lots, and ground mounted systems in urban environment, 2170 GW of distributed PV is installed, which is more than half of the total PV capacity. The presence of heat pumps and battery electric vehicles on the distribution grid level within the system helps eliminate the need for home batteries. 

To conclude, distributed PV, although being more expensive than utility PV, help decreasing total system cost for the energy system. This cost reduction is mainly driven by a diminished requirement for distribution capacity and allows for increased adoption of electric vehicles and electrification of domestic heating. Accurate modeling of distribution power losses and costs significantly influences the achieved cost reduction. Therefore, it is important to separately model distributed and utility PV when examining future energy scenarios. Given the substantial potential of distributed PV, models that trade off computational feasibility with the modelling of distributed resources should be further developed.

\section{Acknowledgments}
P.R. and M.V. partially funded by the AURORA project supported from the European Union’s Horizon 2020 research and innovation programme under grant agreement No. 101036418.

Figure 1 of this study has been designed using images made by Iconjam (utility battery and power to heat icons) and freepik (all other icons) from flaticon.com.

\subsection{Declaration of interests}
The authors declare no competing interests.

\subsection{Author contributions}

Conceptualization: P.R. and M.V.; Software: P.R., E.Z., and M.V.; Investigation: P.R. and M.V.; Methodology: P.R. and M.V.; Project  Supervision: M.V.; Visualization: P.R.; Writing - original draft: P.R.; Writing - review and editing: P.R., M.V., E.Z., and C.G. .

\subsection*{Data and code availability}

The model is implemented by the open energy modeling framework PyPSA and makes use of the the model PyPSA-Eur-Sec v0.6.0 \cite{pypsa_docs} and the costs and technology assumptions included in the technology-data v0.4.0 \cite{pypsa_costs}. Scripts to reproduce the figures included in this paper can be found in \href{https://github.com/Parisra/Distributed-PV-paper}{/github.scripts}. Dataset of the energy system network files are  publicly available at: \href{}{/zenodo.TBD}.


\newpage
\addcontentsline{toc}{section}{References}
\bibliography{Bibliography.bib}

\clearpage
\onecolumn

\glsaddallunused
\printglossary[type=\acronymtype]

\newgeometry{top=25mm}

\beginsupplement
\input{Supplemental}

\restoregeometry
\newpage
\fontsize{10}{12}\selectfont
\addcontentsline{toc}{section}{Supplementary References}
\bibliographystyleS{elsarticle-num}
\bibliographyS{Bibliography.bib}

\clearpage
\tableofcontents

\end{document}

%% file: Supplemental.tex

\clearpage
\addcontentsline{toc}{section}{Supplemental Information}
\section*{Supplementary Information}

\subsection*{S1. Pypsa model and power flow optimization}

A brief explanation on the PyPSA model is provided here.
 Comprehensive information about the optimization method, the objective function, and the constraints implemented in the model can be found in previous studies \citeS{neumann_2022assessments, brown2017, horsch2018pypsa} as well as the model documentation \citeS{pypsadocs}. The main goal of the optimization is to minimise the total annualized system costs. These costs include investment costs and operation and maintenance costs for all system components, as shown in Eq.(1), where \(c_{*}\) is capital cost of the component, \(o_{*}\) is operating cost of the component, \(G_{i,r}\) is generator capacity of technology \(r\) at location \(i\), \(E_{i,s}\) is energy capacity of storage \(s\) at location \(i\), \(P_{l}\) is transmission line capacity for line \(l\), \(F_{k}\) is power capacity of technology \(k\) for conversion and transportation of energy, \(g_{i,r,t}\) is generator dispatch of technology \(r\)  at time \(t\), and \(f_{k,t}\) is dispatch of technology \(k\) at time \(t\). Each time snapshot \(t\) is weighted by the time-step \(w_{t}\), and the sum of time-steps is one year.
\begin{equation*}
\min_{G,F,E,P,g,f} = \left[ \sum_{i,r}^{}c_{i,r}.G_{i,r} + \sum_{k}^{}c_{k}.F_{k}+ \sum_{i,s}^{}c_{i,s}.E_{i,s}+ 
\sum_{l}^{}c_{l}.P_{l}+ 
 \right.
\end{equation*}
\begin{equation}
\left.\sum_{i,r,t}^{}w_{t}\left(\sum_{i,r}^{}o_{i,r}.g_{i,r,t}+\sum_{k}^{}o_{k}.f_{k,t}  \right)   \right]
\end{equation}

A set of constraints are also added to the optimization problem. One of the constraints is that demand is inelastic and must therefore be met completely at each time-step. Other constraints represent different physical and societal limitations such as the maximum renewable potential in every node, maximum transmission expansion, available renewable and non-renewable resources, maximum storage discharge and charge dispatch, and maximum carbon emissions. The objective function and all the constraints are linear, which leads to a linear programme (LP). 
The power flow in the network goes through two main elements: transmission network, and distribution grid. 
Transmission network is comprised of High Voltage Alternating Current (HVAC) lines connecting high-voltage (HV) buses together, as shown in the simplified example in Figure S1. Distribution grid is represented as a single bidirectional connections between each HV bus to its corresponding low-voltage (LV) bus in the same node. 
The simplest way to model power flow is to use the transport model, also known as network flow model. In this case we ignore all physical features of the power transmission such as line resistance and impedance, and only enforce ‘conservation of power’. This is done through the nodal power balance constraint that is modelled with Kirchhoff’s Current Law (KCL), as shown Eq.(2). The KCL constraint ensures that the total inflow power at each bus is equal to the total outflow power plus consumed power.
\begin{equation}
p_{i} = \sum_{l}^{} K_{il}.p_{l}   \quad\quad   \forall i \in \mathcal{N}
\end{equation}
where \(p_{i}\) is the active power injected or consumed at node \(i\), and \(K\) is the incidence matrix of the network graph which summarizes all connections between other nodes and node \(i\) as: not connected (0), connected with start at \(i\) (+1), and connected with end at \(i\) (-1). The power flowing through every line \(p_{l,t}\) is limited by the capacity of the line \(P_{l}\) as shown in Eq.(3), a capacity which is co-optimized if transmission expansion is allowed, as shown in Eq.(4).
\begin{equation}
\left| p_{l,t}  \right|\le \bar{p_{l}}P_{l} \quad\quad   \forall l,t
\end{equation}
The inclusion of  \(\bar{p_{l}}\) as an extra per-unit security margin on the line capacity serves to provide a buffer to account for potential failures of individual circuits (as per the N-1 criterion) and reactive power flows. 
\begin{equation}
\sum_{l}^{} l_{l}.P_{l}\le CAP_{LV} \quad\leftrightarrow \quad\mu_{LV}
\end{equation}
where the sum of transmission capacities \(P_{l}\) multiplied by the lengths \(l_{l}\) is bounded by a transmission volume cap of \(CAP_{LV}\). The Lagrange/KKT multiplier \(\mu_{LV}\) represents the shadow price of a marginal increase in transmission volume. As discussed in previous studies \citeS{neumann_2022assessments}, the transport model does not have enough constraints to produce a unique solution, and this results in arbitrary flows in the network because it does not cost the model anything to transmit power. It also means that the model does not capture possible bottlenecks in the network.

A better representation of the physical features of the power network, while keeping the problem linear, can be obtained with linearised optimal power flow (LOPF) equations, also known as DC power flow. This model adds linear constraints for Kirchhoff’s Voltage Law (KVL) to the KCL constraint of the transport model. The KVL constraint imposes the physical relation between the voltage differences at the extreme of a transmission line and the power flowing through it. The main assumptions behind LOPF are that power flows primarily accordingly to angle differences, no significant voltage shift occurs between the nodes, and line resistances are negligible compared to line reactance\citeS{neumann_2022assessments}. Along with some other simplifying assumptions, the KVL constraint can be linearised to calculate the power flow \(p_{i}\) in each line with Eq.(3):
\begin{equation}
p_{l} = \frac{\Theta_{i}-\Theta_{j}}{x_{l}}
\end{equation}
where \(\Theta_{i}\) and \(\Theta_{j}\) are voltage angles at nodes \(i\) and \(j\), and \(x_{l}\) is the line reactance. In the model, KVL is imposed by means of the cycle matrix where \(C_{lc}\) contains information on which line \(l\) is an element of a closed loop \(c\) \citeS{neumann_2022assessments}.
\begin{equation}
\sum_{l}^{} C_{lc}P_{l}x_{l}=0 \quad \forall c\in C
\end{equation}
Although LOPF improves upon the transport model by including reactance and voltage angles into the model, it also assumes negligible resistance, so power flow in the transmission network in our model is effectively lossless.
\begin{figure}[H]
\renewcommand*{\thefigure}{S\arabic{figure}}
\includegraphics[width=0.8\textwidth,center]{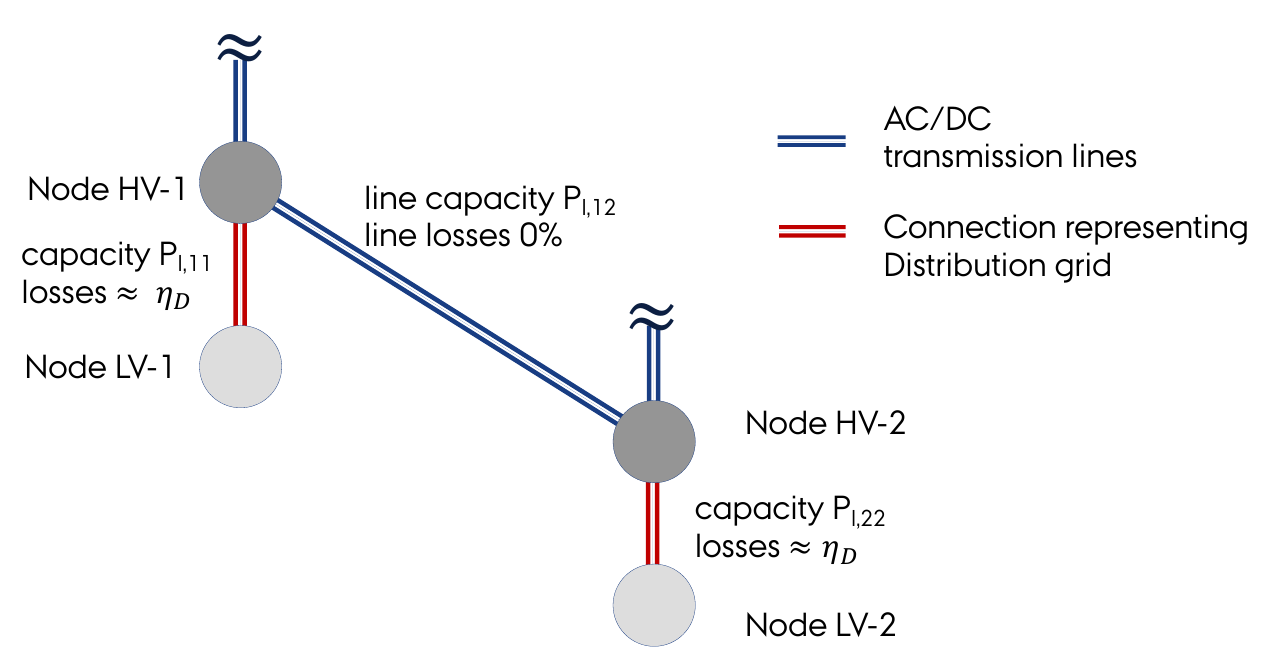}
\caption{Schematic of connections between two high-voltage and low-voltage nodes in the network. Transmission lines are modeled using LOPF equations and line capacity \(P_{l}\) is optimized assuming zero resistive losses. Distribution grid connections are modeled using the Transport model and  capacity \(P_{l}\) is optimized assuming a constant efficiency of \(\eta_{D}\). }
\end{figure}
Modeling the network with LOPF is significantly more demanding computationally than using the transport model. To minimise computational complexity, only the transmission network, meaning all AC and DC HV transmission lines between HV buses, are modelled with LOPF. The assumptions behind LOPF are also more reasonable when looking at transmission lines. The distribution grid is modelled as a single connections between the HV bus and LV bus for each node with the transport model, so the only constraint that applies to it is the KCL constraint. To represent power flow losses in the distribution network, a constant efficiency is assumed for all distribution connections, which could be 100\% or 90\% (10\% power losses) as mentioned in the text.

\subsection*{S2. Distribution grid modeling}

Some small-scale studies consider several levels for the grid with a high-high voltage level for transmission network and lower levels (high, medium, low) for distribution grid \citeS{muller2019}. In our model we consider only one HV level and one LV level per node. Another approach to modelling the distribution grid is to use 2 unidirectional connections instead of one bidirectional connection\citeS{clack2020}. This was tested for the scenario B under different transmission and distribution grid assumptions, as shown in Figure S2. The results showed no significant difference for system capacity mix. This is due to the fact that reverse flows in the distribution grid are very low and the system does not consider it cost-efficient to install any capacity for this direction. Hence, we follow a simple approach using one bidirectional connection.

\begin{figure}[H]
\renewcommand*{\thefigure}{S\arabic{figure}}
\includegraphics[width=1\textwidth,]{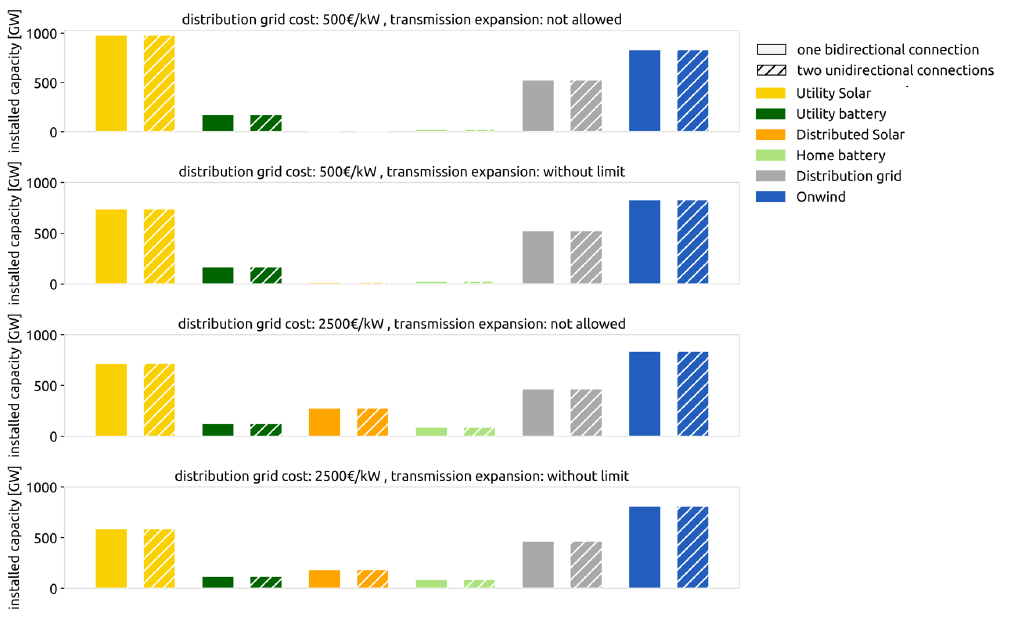}
\caption{Comparison of installed capacity of major technologies for the power sector under different transmission grid allowance and distribution grid cost assumptions when the distribution grid is modeled as one bidirectional connection vs. as two unidirectional connections. }
\end{figure}

A number of studies including grid reinforcement costs are shown in Table S1. Some of these studies represent grid costs as \texteuro /m for distribution lines, or as \texteuro /kWh for the amount of energy transfer through the distribution grid. To convert these costs to the unit \texteuro /kW, the following is assumed for the European distribution grid : 300 million customers, 10 millions km of power lines, 2800 annual TWh demand \citeS{rullaud2020distribution} and 542 GW peak hourly demand \citeS{ENTSOE_2017}

Let us assume that we have a cost of 1 \texteuro /km for distribution grid. We can multiply this by the number 10 million (km), which would result in 10 million \texteuro  as the total cost of the distribution grids in Europe. We can then divide this number by the European energy system peak load, which is 542 GW, and the resulting number could be considered the cost of the distribution grid in \texteuro /kW. This methodology is rather crude, so only the studies where the numbers fall within an established criterion are presented here.

\begin{table}[H]
\renewcommand*{\thetable}{S\arabic{table}}
\centering
\caption{An overview of distribution grid reinforcement costs from selected studies}
\scriptsize
\begin{tabular}{p{3cm}p{3.2cm}p{1.5cm}p{1.5cm}p{1.5cm}p{1.5cm}}
\toprule
References & Capital cost                  (\texteuro /kW) & Lifetime (years) & Discount rate \\
\hline
Meunier et al. (2021) \citeS{meunier2021cost} & 15- 32 (from   per dwelling) & 33 & - \\
Miller et. All (2022) \citeS{miller2022grid} & 110 (from line costs) & - & - \\
Allard et al. (2020) \citeS{allard2020considering} & 148-232  (from line costs) & - & 8\% \\
Study by Imperial College, NERA and DNV GL (2014) \citeS{european2014integration} & 75-270  (from line costs) & - & 10\% \\
Gupta et al. (2021) \citeS{gupta_2021} & 51-220,1385,143 (reinforcement for PV, HP, EV) & 40-50 & 3.80\% \\
Lumbreras et. Al (2018) \citeS{lumbreras2018} & 0-80 (reinforcement for PV) & - & -  \\
\bottomrule
\end{tabular}
\end{table}

\subsection*{S3. Sensitivity of system capacity mix to spatial resolution}

\begin{figure}[H]
\renewcommand*{\thefigure}{S\arabic{figure}}
\includegraphics[width=0.8\textwidth,center]{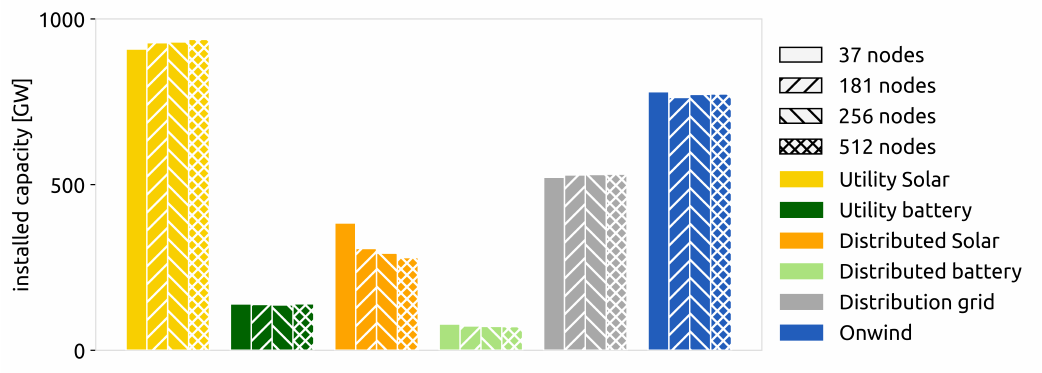}
\caption{Comparison of installed capacity of major technologies for the scenario B with distributed generation and storage under different spatial resolutions with 37, 181, 256, and 512 nodes}

\end{figure}

\subsection*{S4. Sensitivity of installed distributed solar capacity to grid costs and losses}

\begin{figure}[H]
\renewcommand*{\thefigure}{S\arabic{figure}}
\includegraphics[width=0.55\textwidth, center]{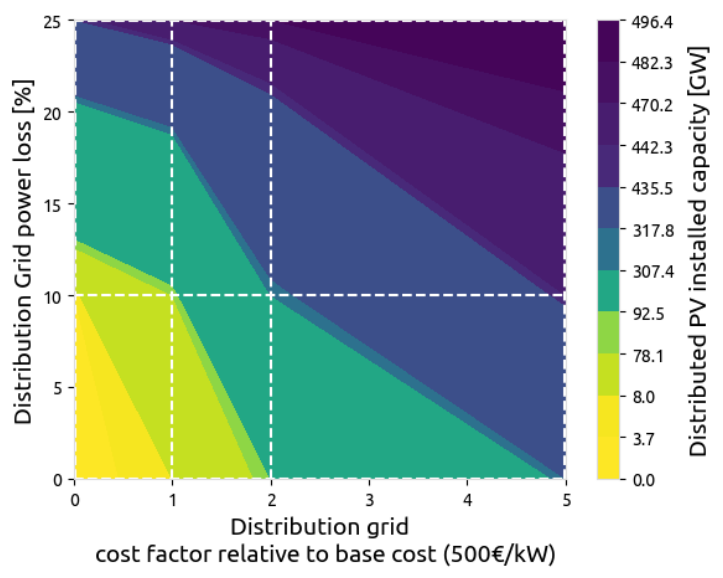}
\caption{Sensitivity of the installed capacity of distributed solar for scenario B to distribution grid costs of 0, 500, and 2500 \texteuro /kW, and distribution power loss of 0, 10\%, 25\%. The difference for grid cost factor of 0.01 and 1 is negligible, so lower costs than base cost (500 \texteuro /kW) are ignored in the analysis.}

\end{figure}

\subsection*{S5. Sensitivity of system cost savings from distributed solar to emissions target}

\begin{figure}[H]
\renewcommand*{\thefigure}{S\arabic{figure}}
\includegraphics[width=0.8\textwidth, center]{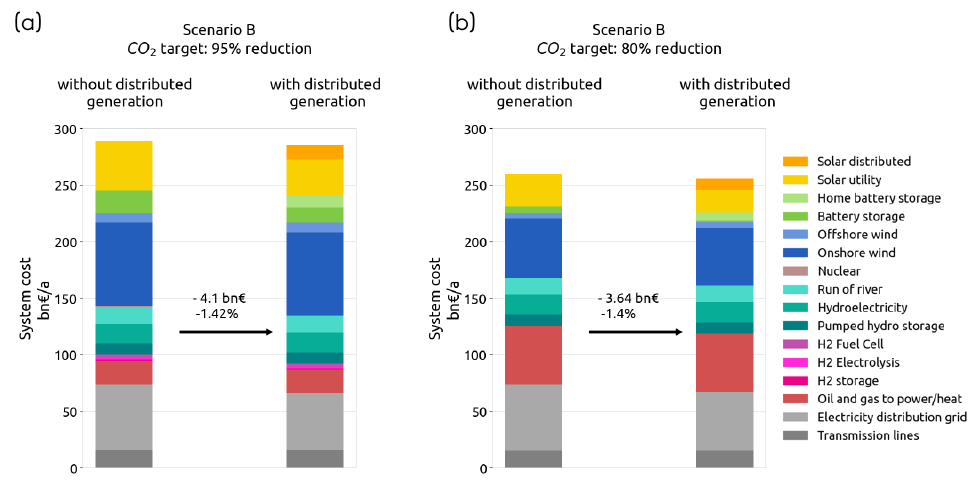}
\caption{Total costs of scenario B with and without distributed technologies for a) 95\% carbon emissions reduction compared to 1990 levels, and b) 80\% carbon emissions reduction compared to 1990 levels. The cost savings and system generation mix remain unchanged, showing that the results are robust for lower emission targets. }
\end{figure}

\subsection*{S6. Changes in installed capacity and costs of each technology for all scenarios}

\begin{figure}[H]
\renewcommand*{\thefigure}{S\arabic{figure}}
\includegraphics[width=0.9\textwidth,]{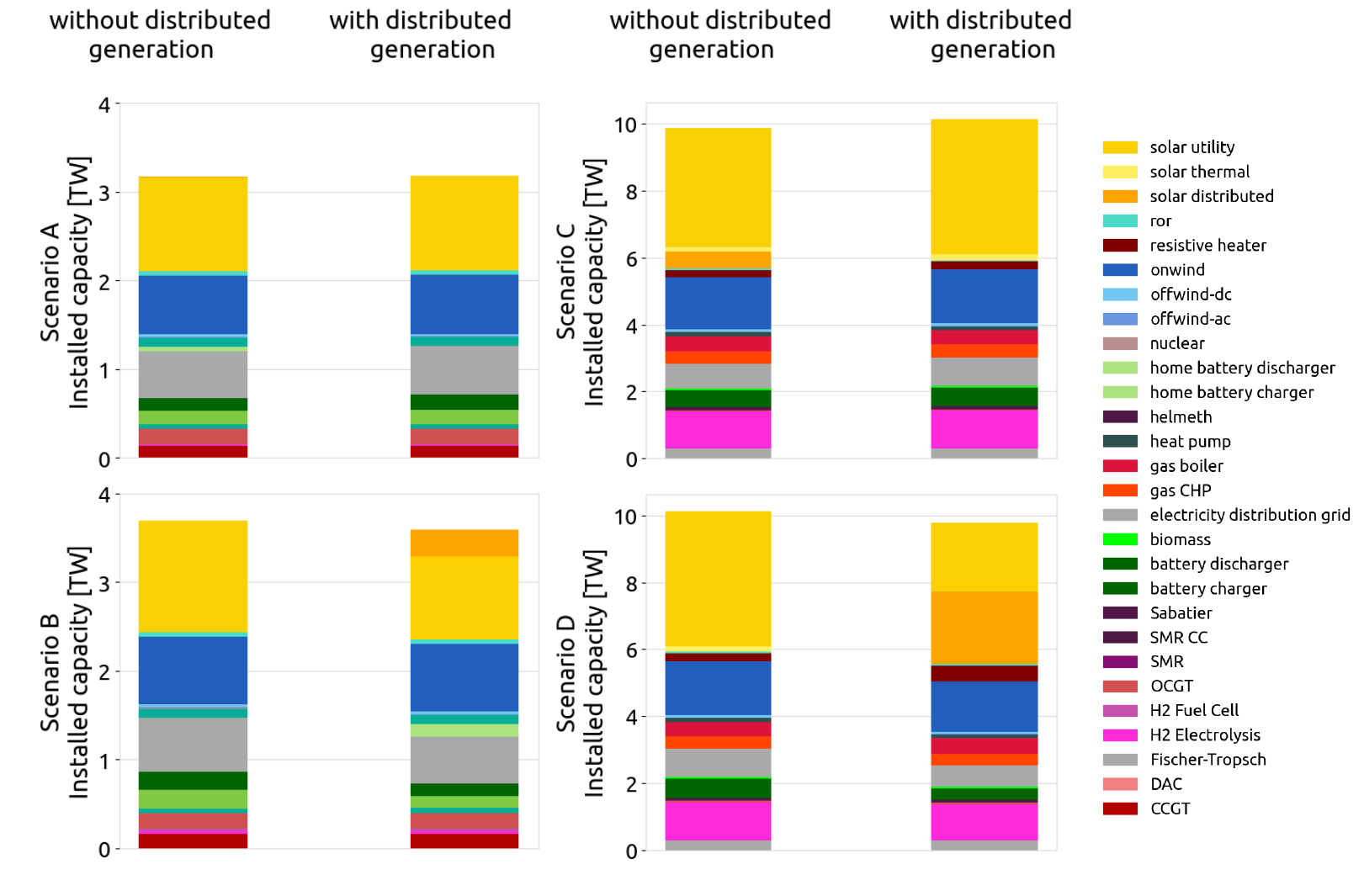}
\caption{System capacity mix for scenarios A, B and C: with and without distributed generation.}
\end{figure}

\begin{figure}[H]
\renewcommand*{\thefigure}{S\arabic{figure}}
\includegraphics[width=0.7\textwidth, center]{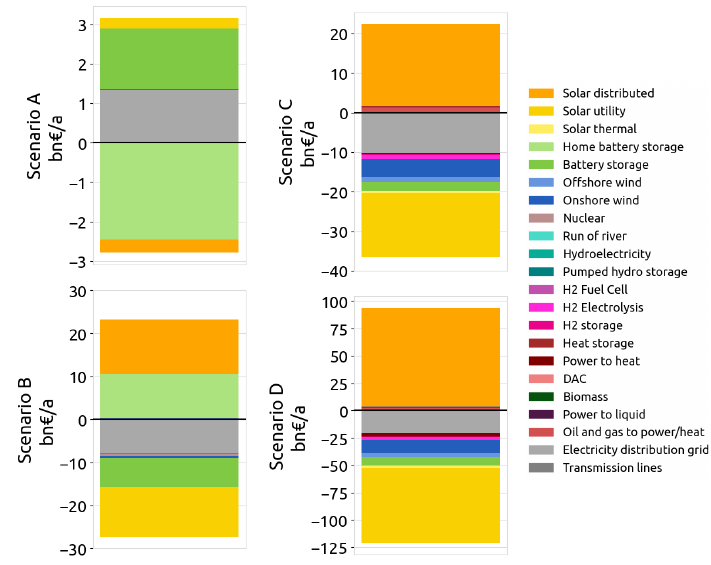}
\caption{Changes in system costs when including distributed solar and home batteries. Distributed solar is the main technology with increasing costs. Solar utility and distribution grid are the main components with decreasing costs.}
\end{figure}

\subsection*{S7. Energy generation mix for all scenarios for Italy, Germany, and the whole system}

\begin{figure}[H]
\renewcommand*{\thefigure}{S\arabic{figure}}
\includegraphics[width=0.8\textwidth, center]{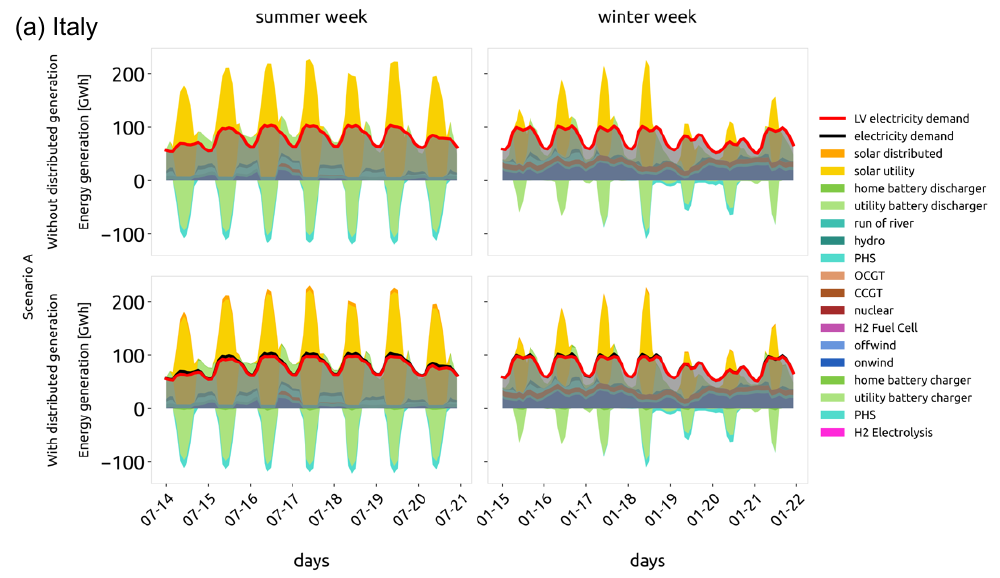}
\includegraphics[width=0.8\textwidth,center]{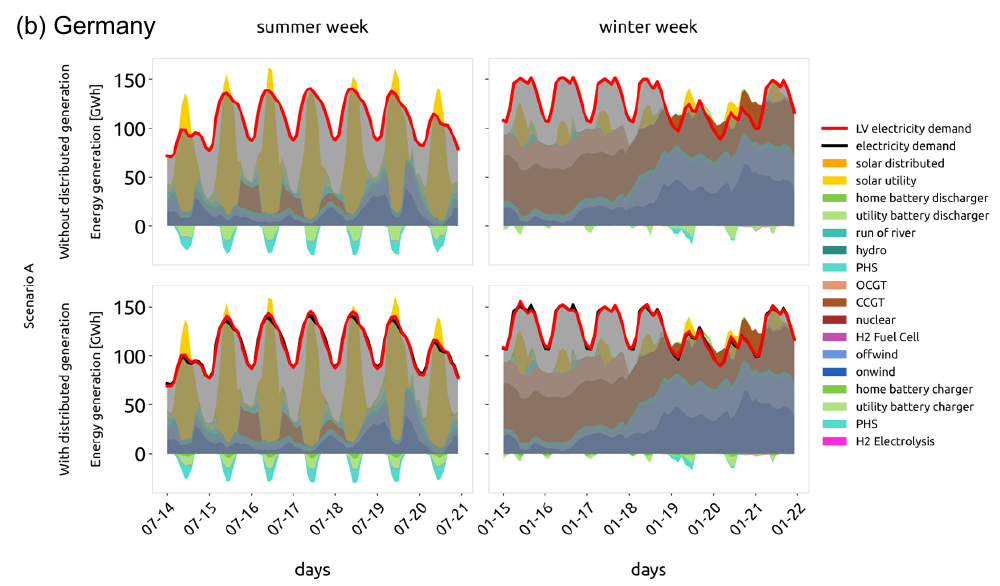}
\includegraphics[width=0.8\textwidth,center]{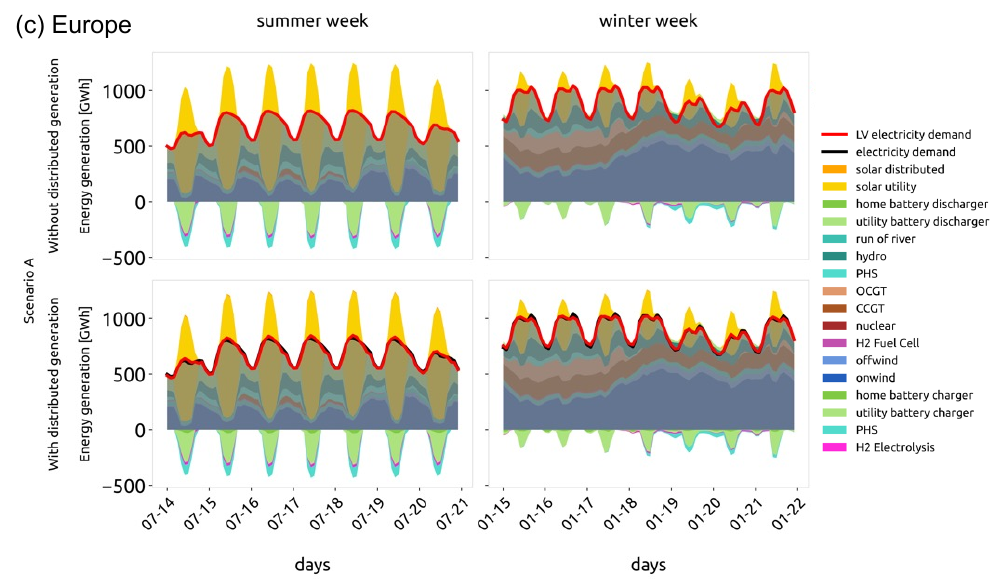}
\caption{Energy generation mix for a week in Summer without(top) and with (bottom) distributed solar and batteries for scenario A}
\end{figure}

\begin{figure}[H]
\renewcommand*{\thefigure}{S\arabic{figure}}

\includegraphics[width=0.8\textwidth,center]
{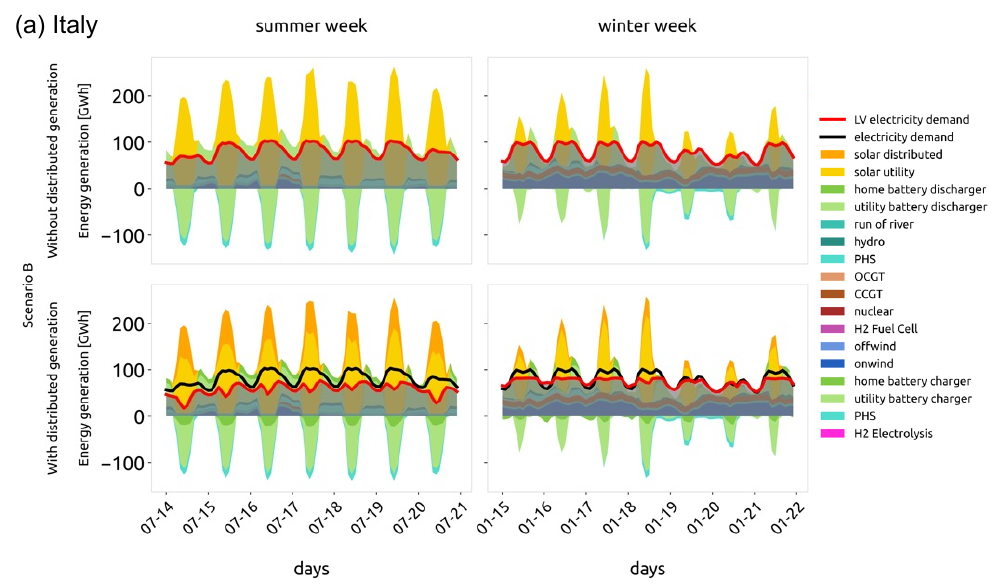}
\includegraphics[width=0.8\textwidth,center]{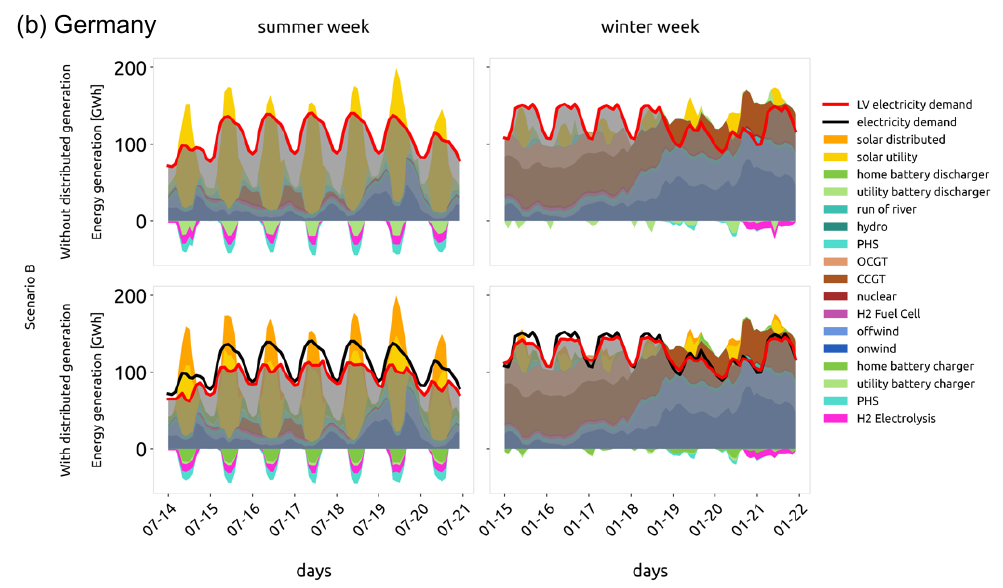}
\includegraphics[width=0.8\textwidth,center]{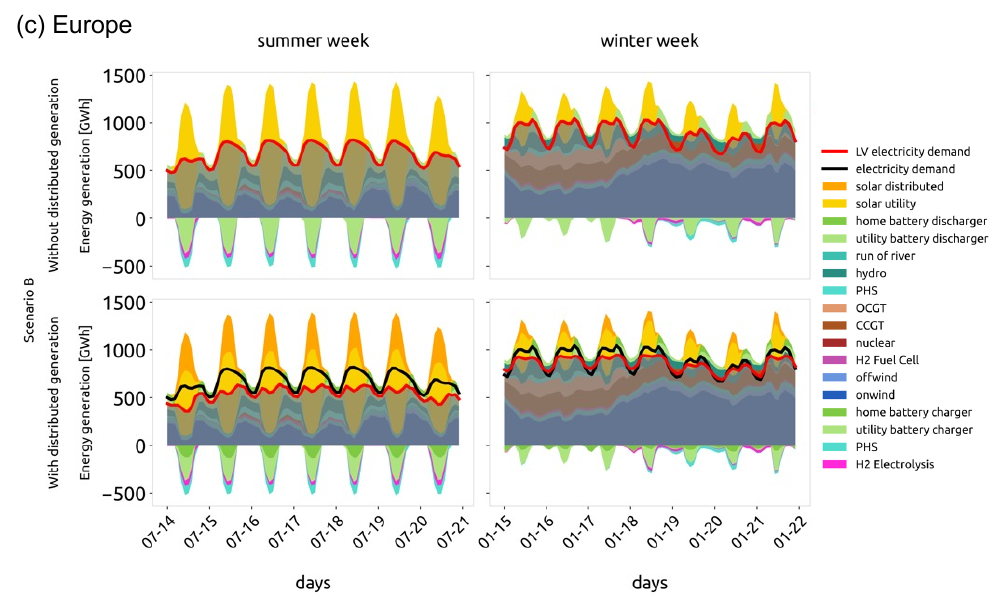}
\caption{Energy generation mix for a week in Summer without(top) and with (bottom) distributed solar and batteries for scenario B}
\end{figure}

\begin{figure}[H]
\renewcommand*{\thefigure}{S\arabic{figure}}

\includegraphics[width=0.8\textwidth,center]{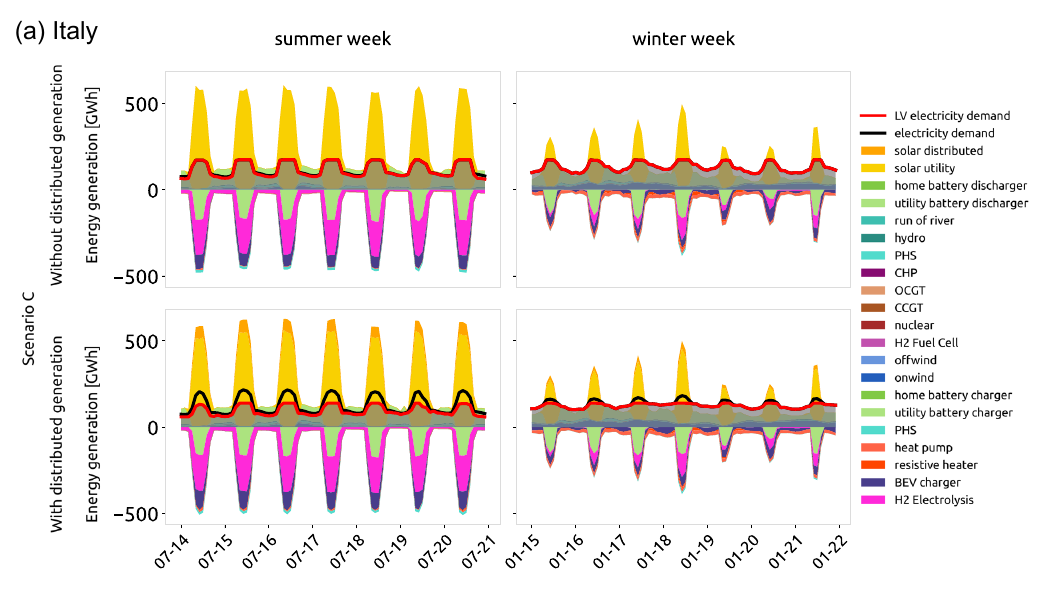}
\includegraphics[width=0.8\textwidth,center]{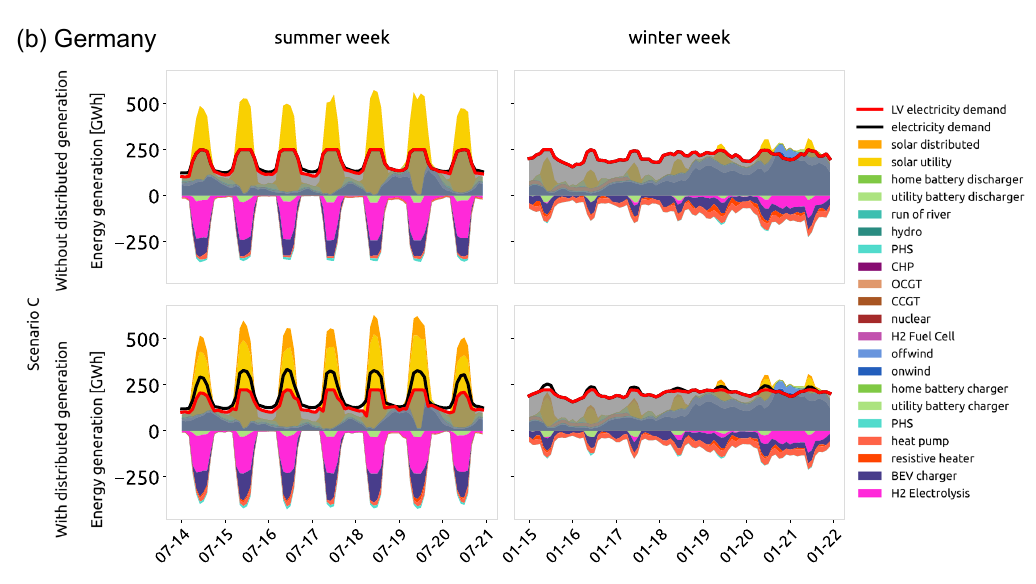}
\includegraphics[width=0.8\textwidth,center]{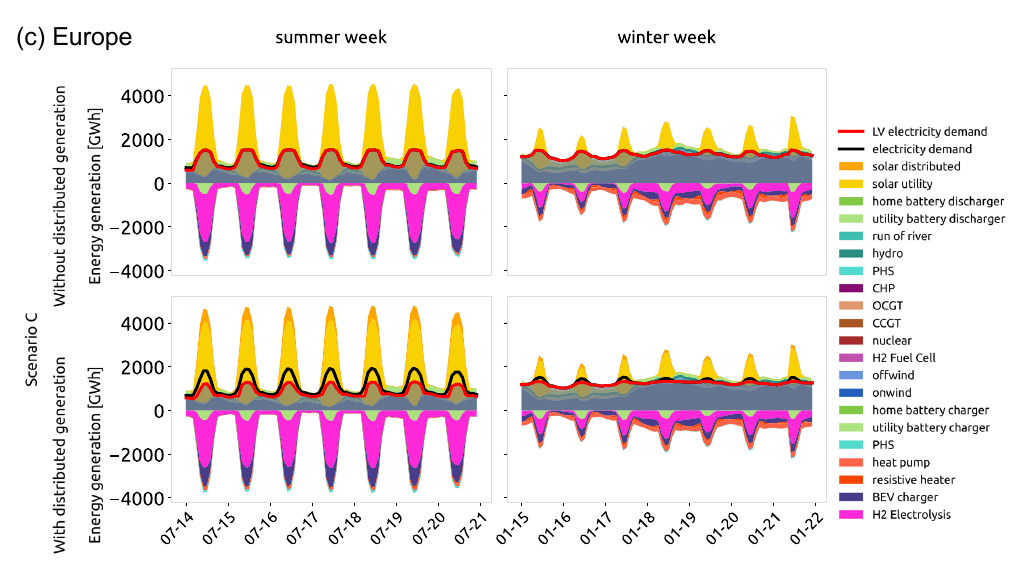}
\caption{Energy generation mix for a week in Summer without(top) and with (bottom) distributed solar and batteries for scenario C}
\end{figure}

\begin{figure}[H]
\renewcommand*{\thefigure}{S\arabic{figure}}

\includegraphics[width=0.55\textwidth,center]{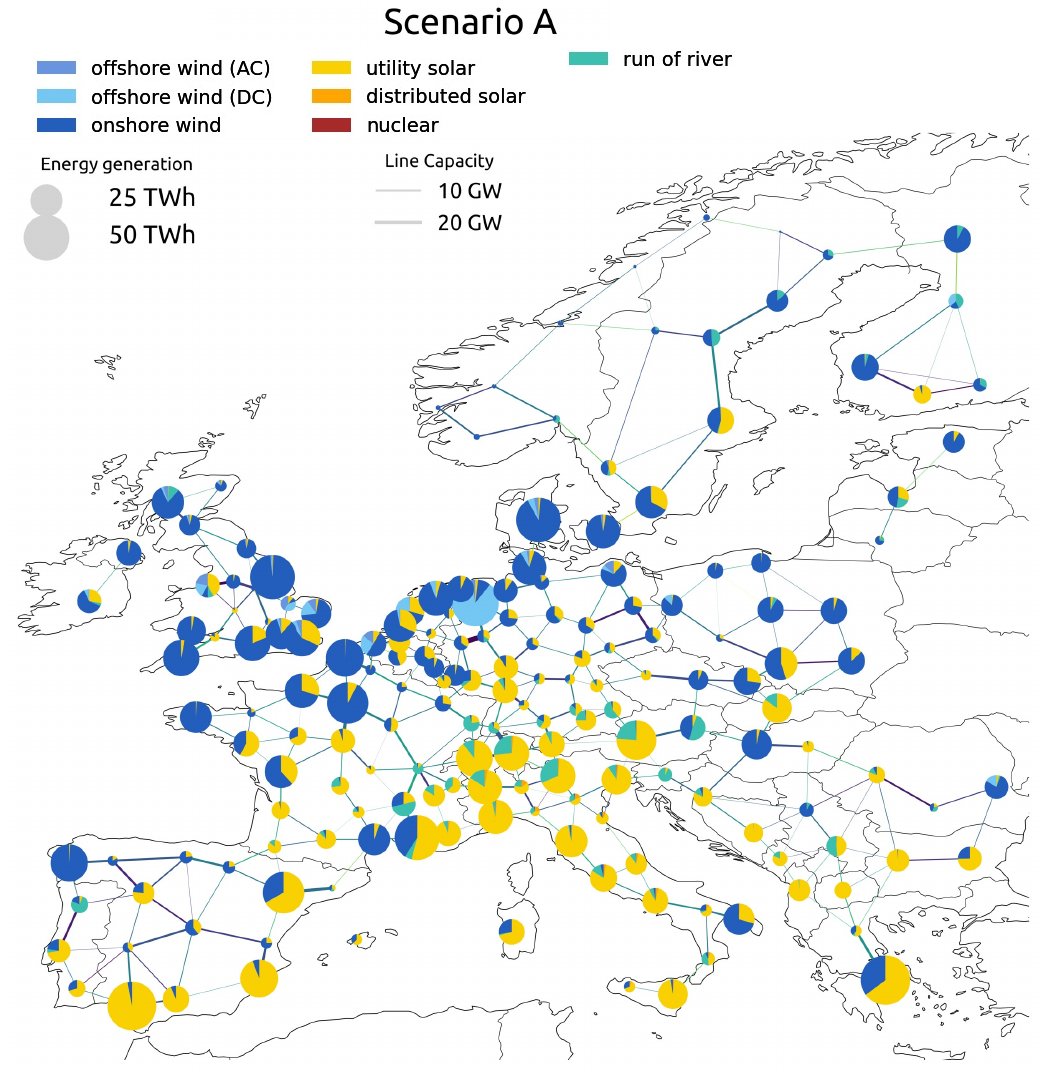}
\caption{Regional map for Scenario A with distributed generation showing share of major technologies in total annual electricity generation}
\end{figure}

\begin{figure}[H]
\renewcommand*{\thefigure}{S\arabic{figure}}

\includegraphics[width=0.55\textwidth,center]{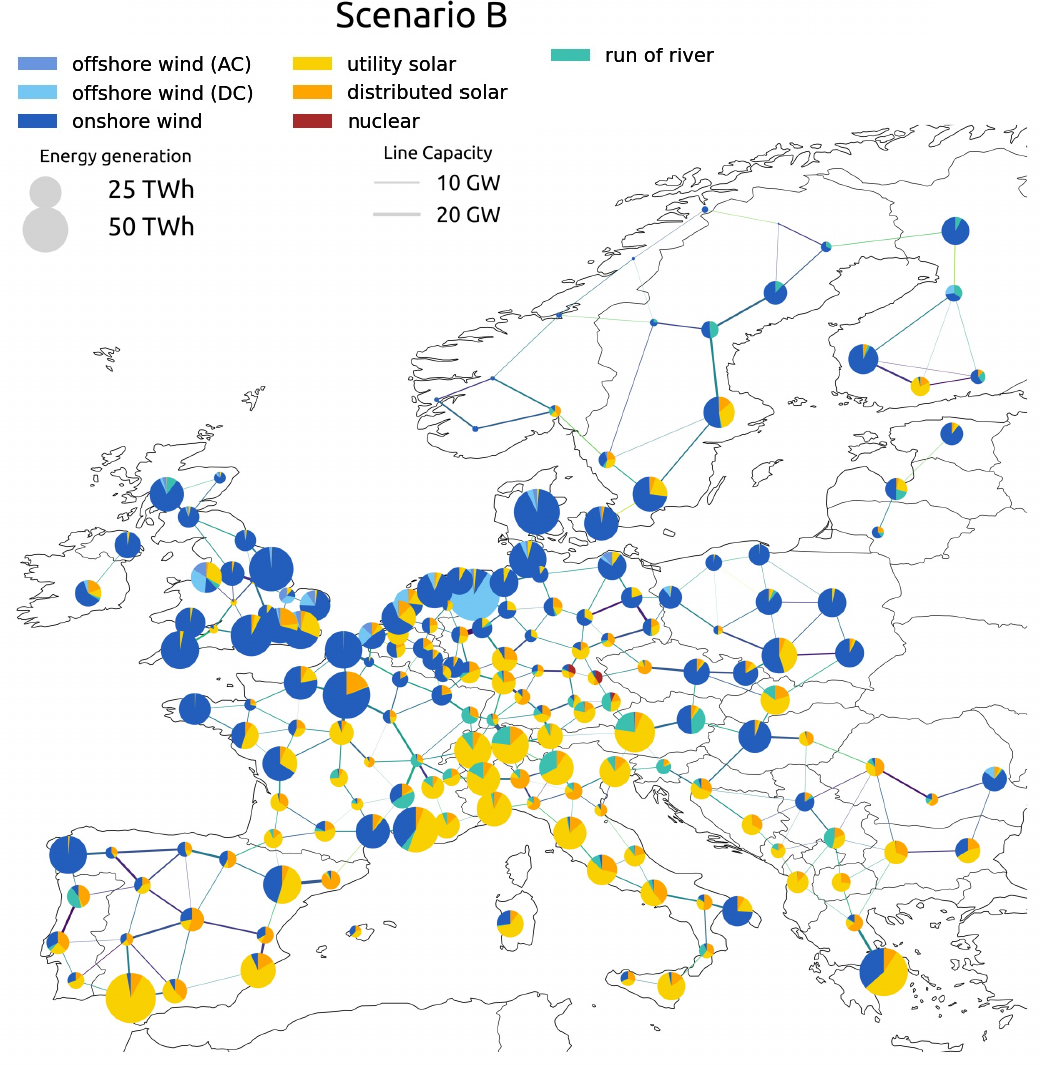}
\caption{Regional map for Scenario B with distributed generation showing share of major technologies in total annual electricity generation}
\end{figure}

\begin{figure}[H]
\renewcommand*{\thefigure}{S\arabic{figure}}

\includegraphics[width=0.55\textwidth,center]{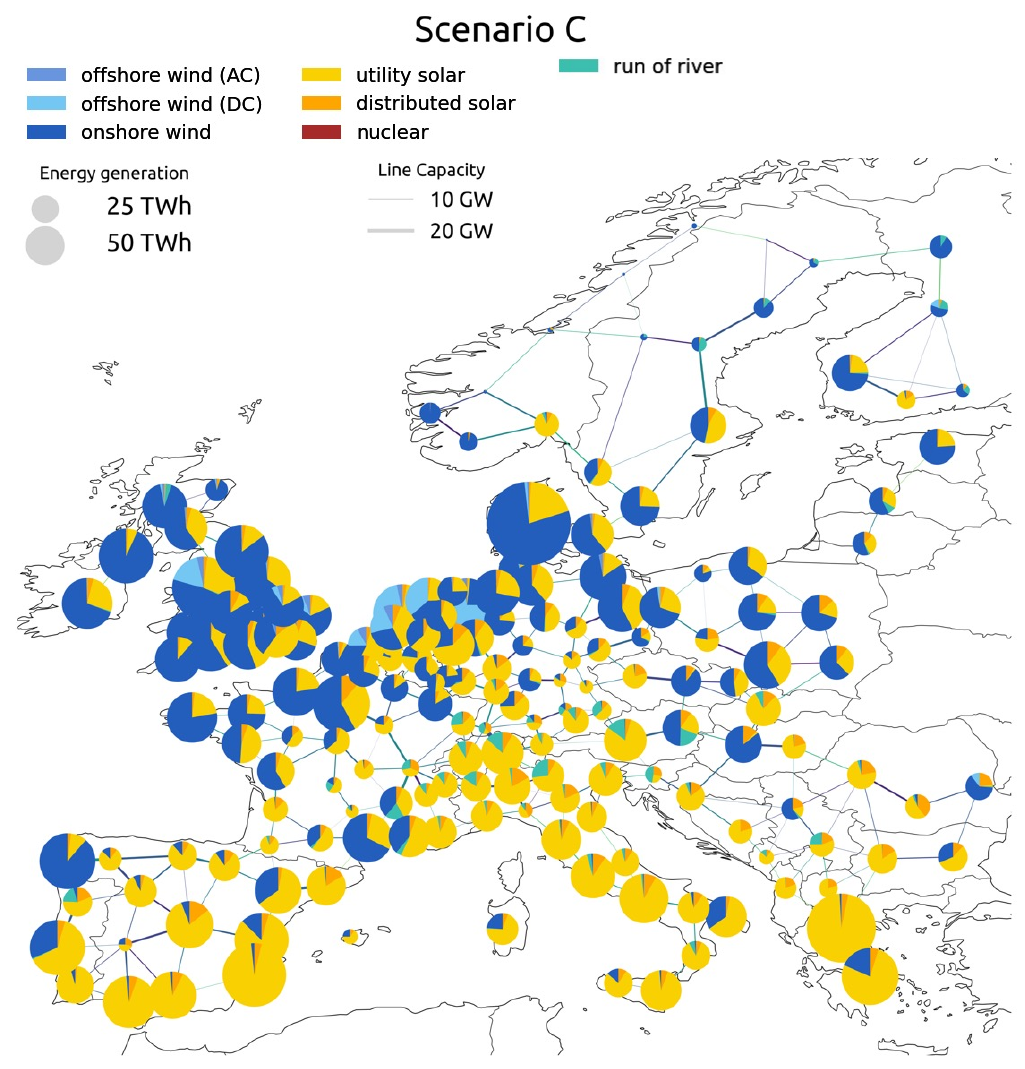}
\caption{Regional map for Scenario C with distributed generation showing share of major technologies in total annual electricity generation}
\end{figure}

\begin{figure}[H]
\renewcommand*{\thefigure}{S\arabic{figure}}

\includegraphics[width=0.55\textwidth,center]{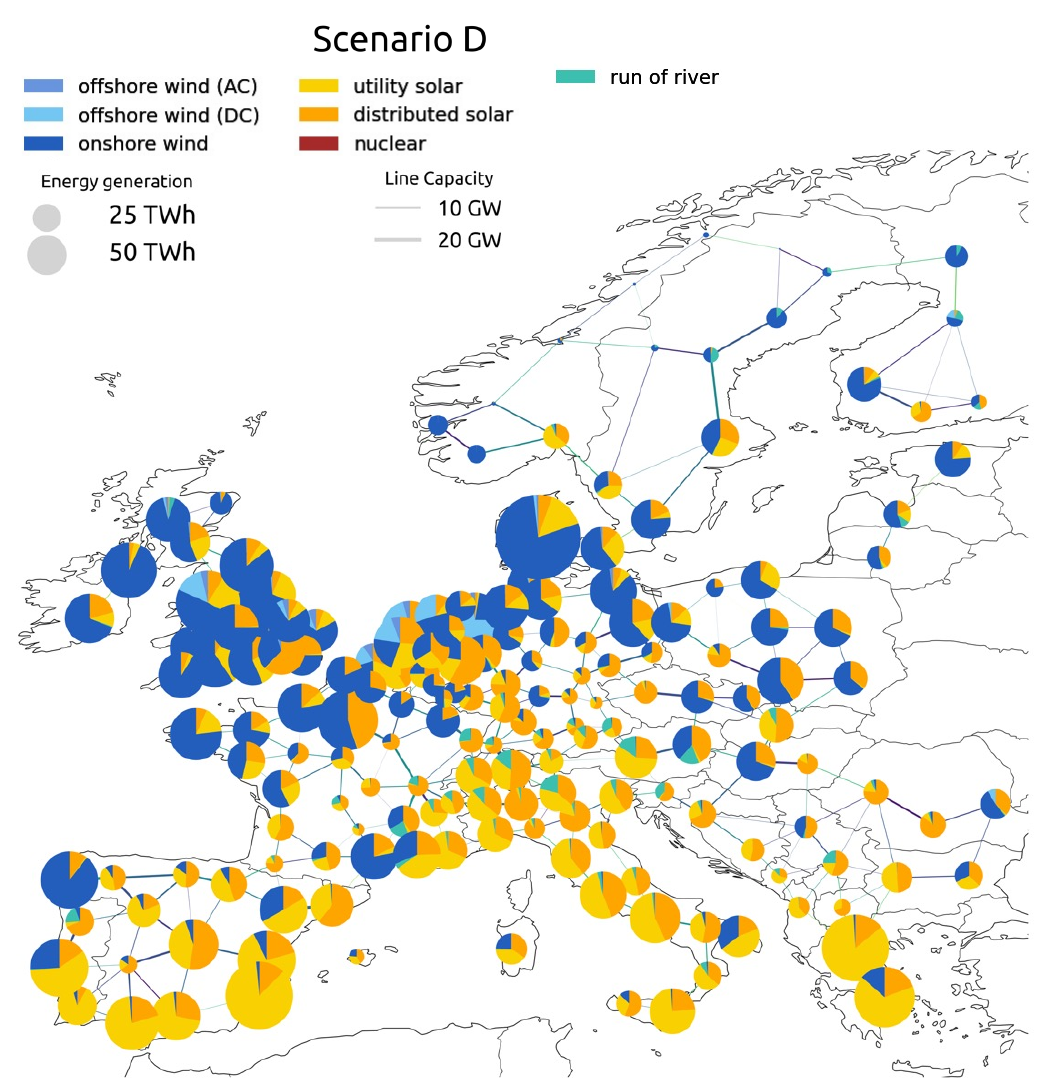}
\caption{Regional map for Scenario D with distributed generation showing share of major technologies in total annual electricity generation}
\end{figure}

\subsection*{S8. Installed capacity of different technologies to maximum installable capacity}

\begin{figure}[H]
\renewcommand*{\thefigure}{S\arabic{figure}}

\includegraphics[width=1\textwidth,]{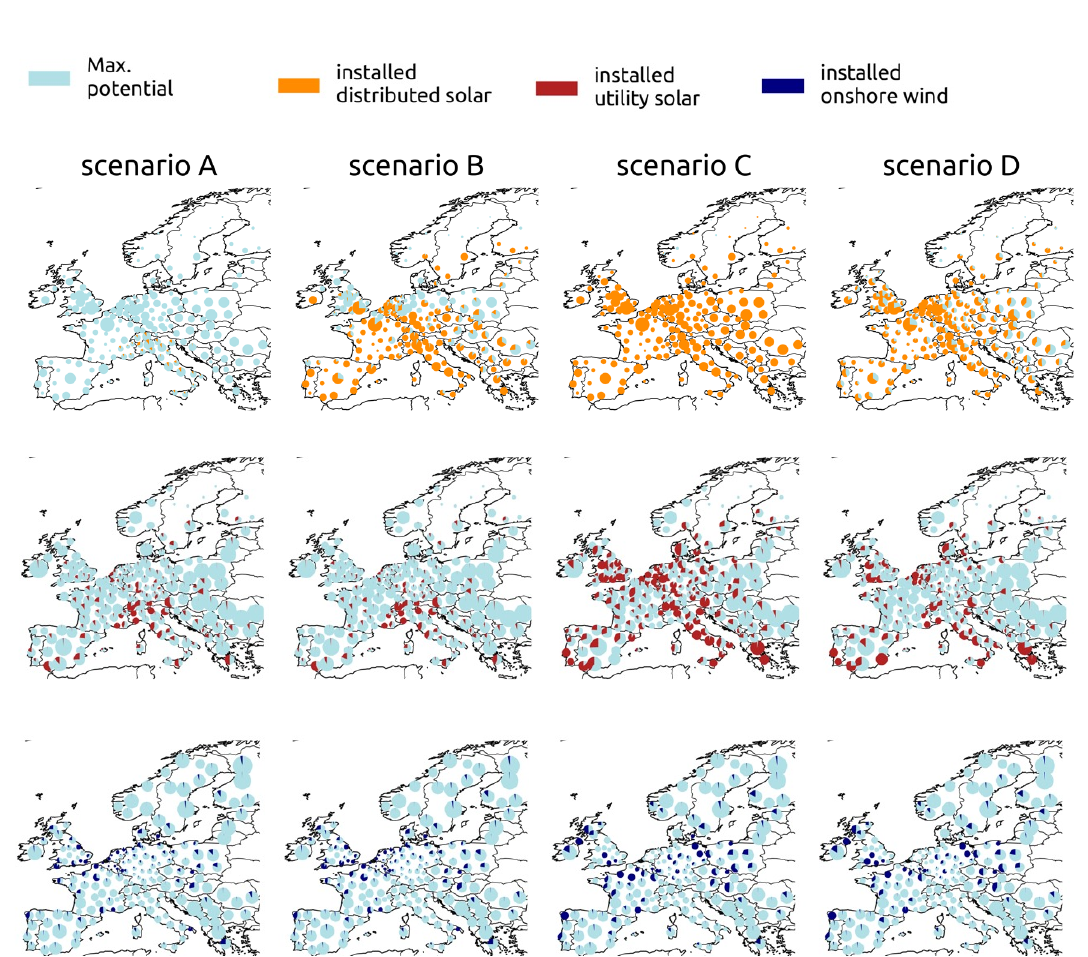}
\caption{Map of installed capacity vs. maximum installable capacity for distributed solar, utility solar, and wind in scenarios A, B, and C. All the capacities increase for the sector-coupled scenario C to meet additional demand. Installed capacity of distributed solar reaches maximum for most nodes in scenario C, and about half the nodes for scenario B}
\end{figure}

\subsection*{S9. Transmission network role}

\begin{figure}[H]
\renewcommand*{\thefigure}{S\arabic{figure}}

\includegraphics[width=1\textwidth,]{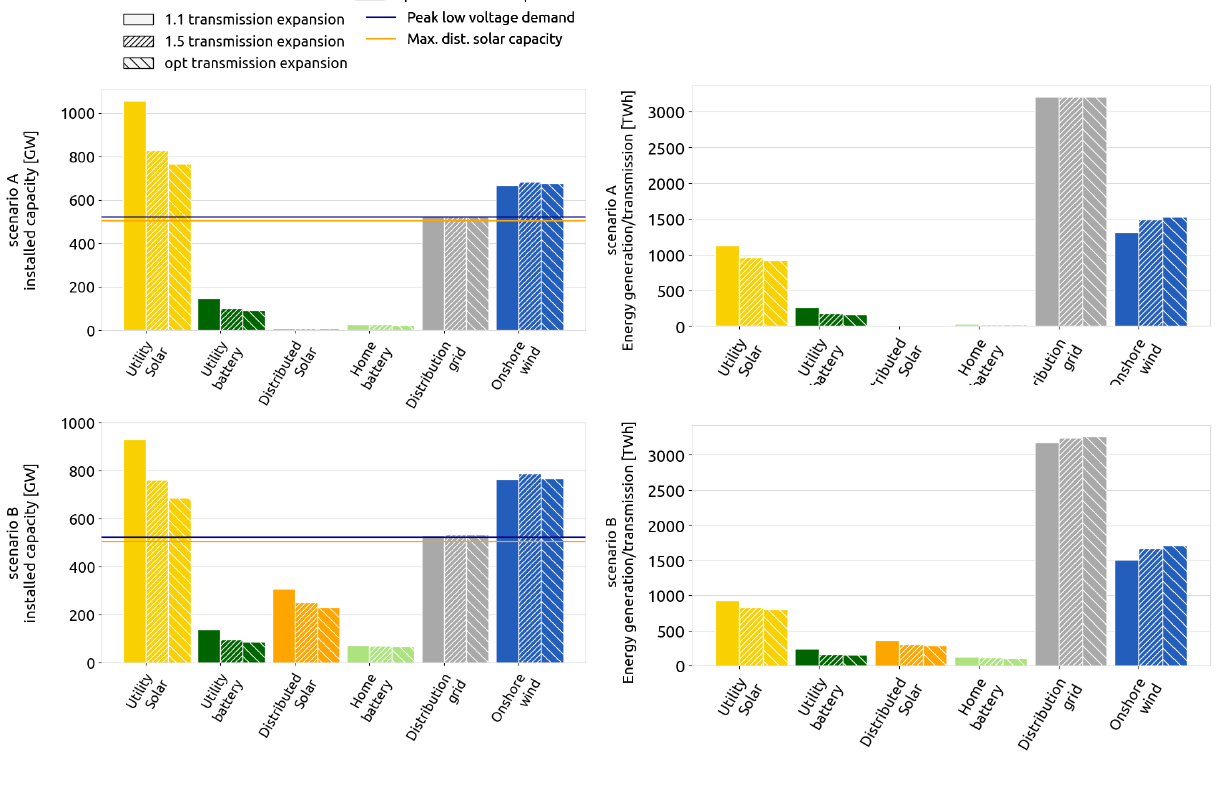}   
\caption{Sensitivity of the installed capacity and the energy generated/transmitted for solar utility, utility batteries, distributed solar, home batteries, and the distribution grid to transmission network expansion allowance for top) scenario A and bottom) scenario B. All figures are for the case when the scenario includes distributed generation and storage. Both scenario A and scenario B show a reduction in installed capacities of utility solar and utility battery when the transmission network allowance goes from 10\% of the current network capacity to 50\%, and to 'no-limit' expansion optimised by the model. Energy generation figures both show that wind generation is increasing for higher transmission network capacity, compensating for the reduction in solar energy. Due to easy transportation of energy from locations with high wind potential, although installed wind capacity decreases for optimal transmission allowances, wind energy generation continues to increase. The transmission network capacity is equal to 1.7 times the original transmission network for 'no-limit' expansion. }
\end{figure}

\begin{figure}[H]
\renewcommand*{\thefigure}{S\arabic{figure}}

\includegraphics[width=0.9\textwidth, center]{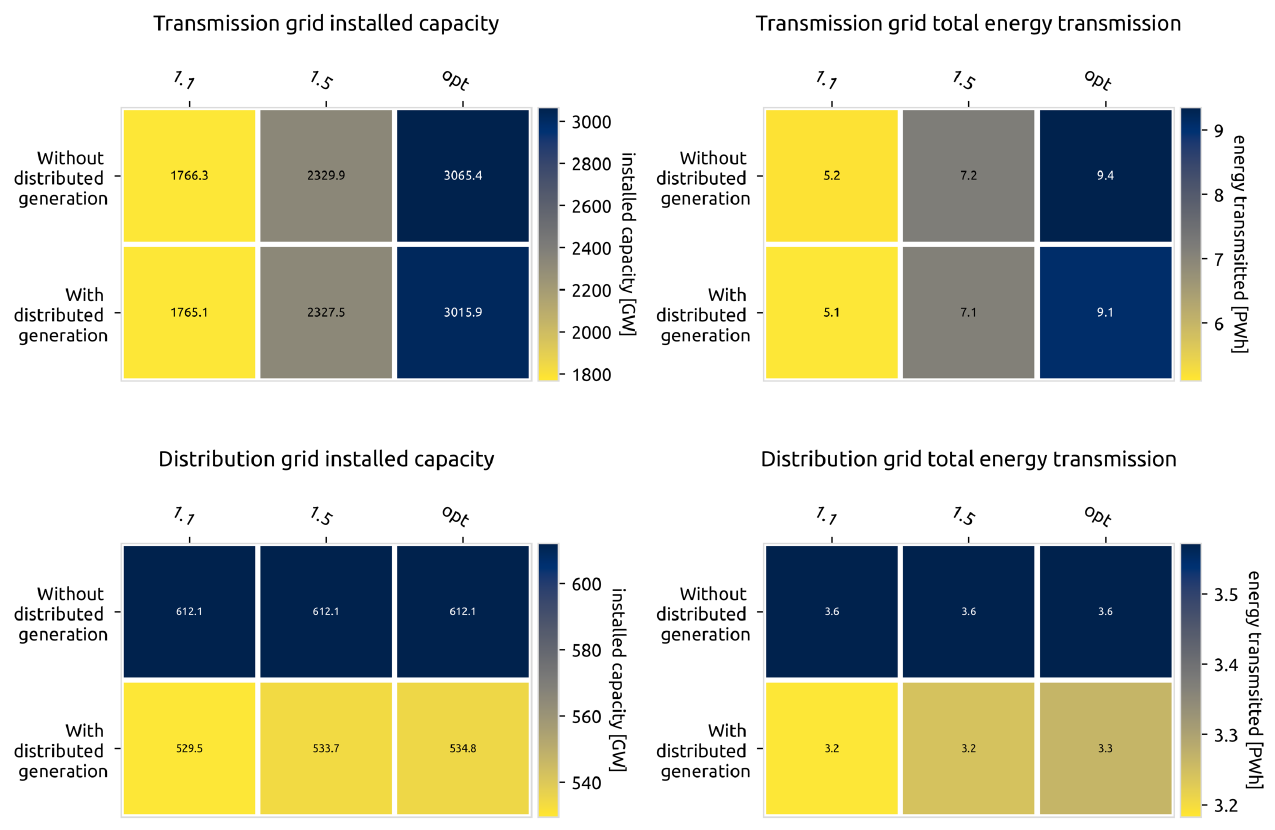}   
\caption{An overall look at the interdependence of transmission network and distribution network for scenario B for different transmission expansion volumes (10\%, 50\%, and no-limit expansion optimised by model). Transmission network's capacity and energy transmission do not show a noticeable change when distributed generation and storage technologies are not available. Distribution network's capacity and energy transmission shows a weak dependency on transmission network capacity, with both capacity and energy transmission increasing as the transmission network is allowed to expand more.}
\end{figure}

\begin{figure}[H]
\renewcommand*{\thefigure}{S\arabic{figure}}

\includegraphics[width=0.9\textwidth, center]{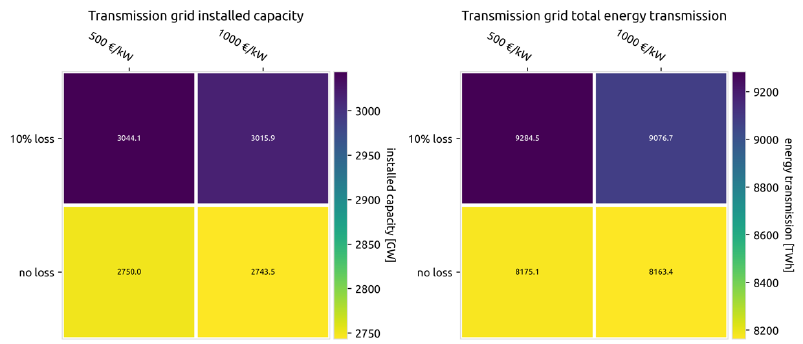}   
\caption{Changes in the installed capacity and total energy transmission through the transmission network for different distribution grid costs and losses. The transmission network is allowed to expand without limit in all cases. The figures show that 1) assumptions on distribution cost have no impact on transmission grid optimal capacity and total transmitted energy, and 2) assumption on distribution losses has a small impact as more energy needs to be transmitted to make up for distribution loss. 
}
\end{figure}

\subsection*{S10. Effect of distributed technologies on electricity transmission, dominant cycle, and price }

\begin{figure}[H]
\renewcommand*{\thefigure}{S\arabic{figure}}

\includegraphics[width=0.75\textwidth, center]{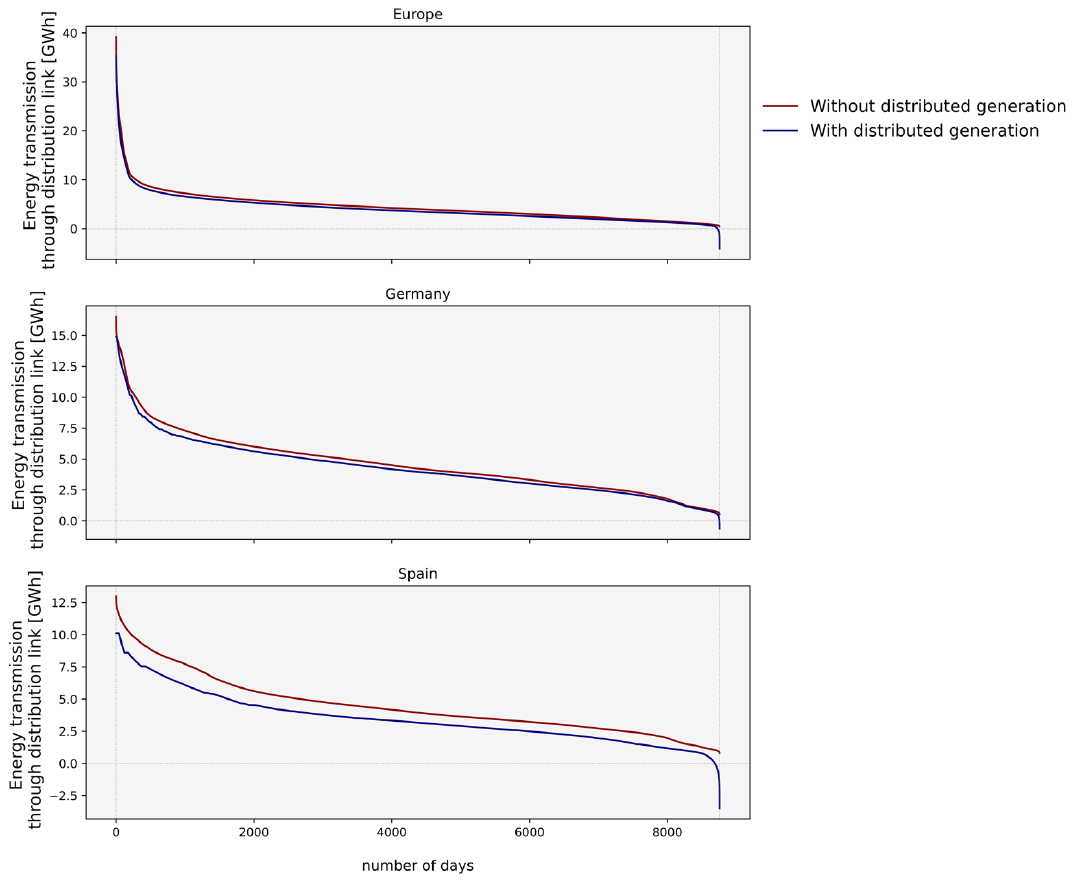}
\caption{Duration curve for energy transfer in the distribution grid for entire system, Germany, and Spain. As expected, the LV demand peak reduction is highest for Spain, which has a higher installed capacity of distributed solar. }
\end{figure}

\begin{figure}[H]
\renewcommand*{\thefigure}{S\arabic{figure}}

\includegraphics[width=1\textwidth,]{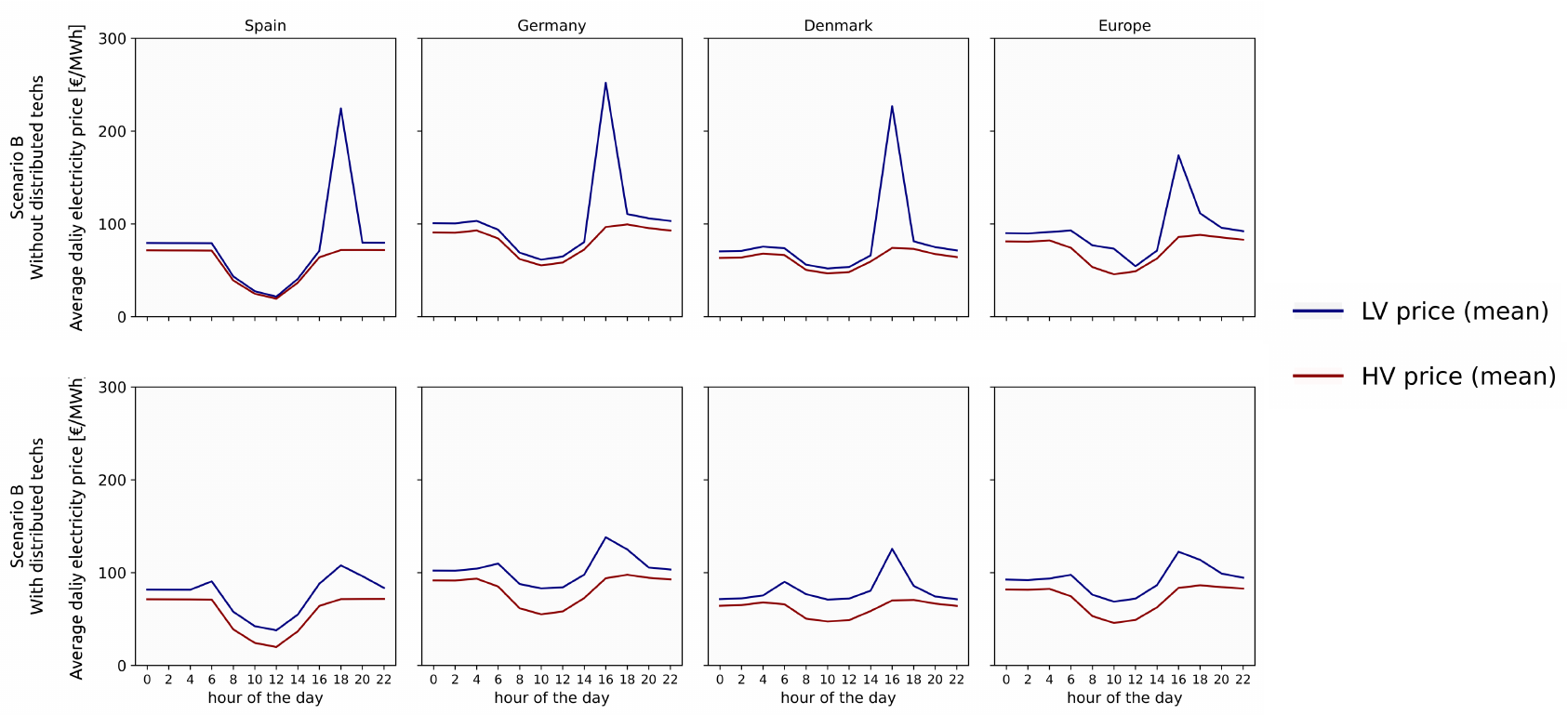}
\caption{Comparison of electricity price for scenario B with and without distributed technologies for different nodes. When distributed generation and storage are available, the electricity price has a lower variation. The low electricity prices in midday (known as duck curve) can be seen for all nodes and happens on both the LV buses and the HV buses.  }
\end{figure}

\begin{figure}[H]
\renewcommand*{\thefigure}{S\arabic{figure}}

\includegraphics[width=1\textwidth,]{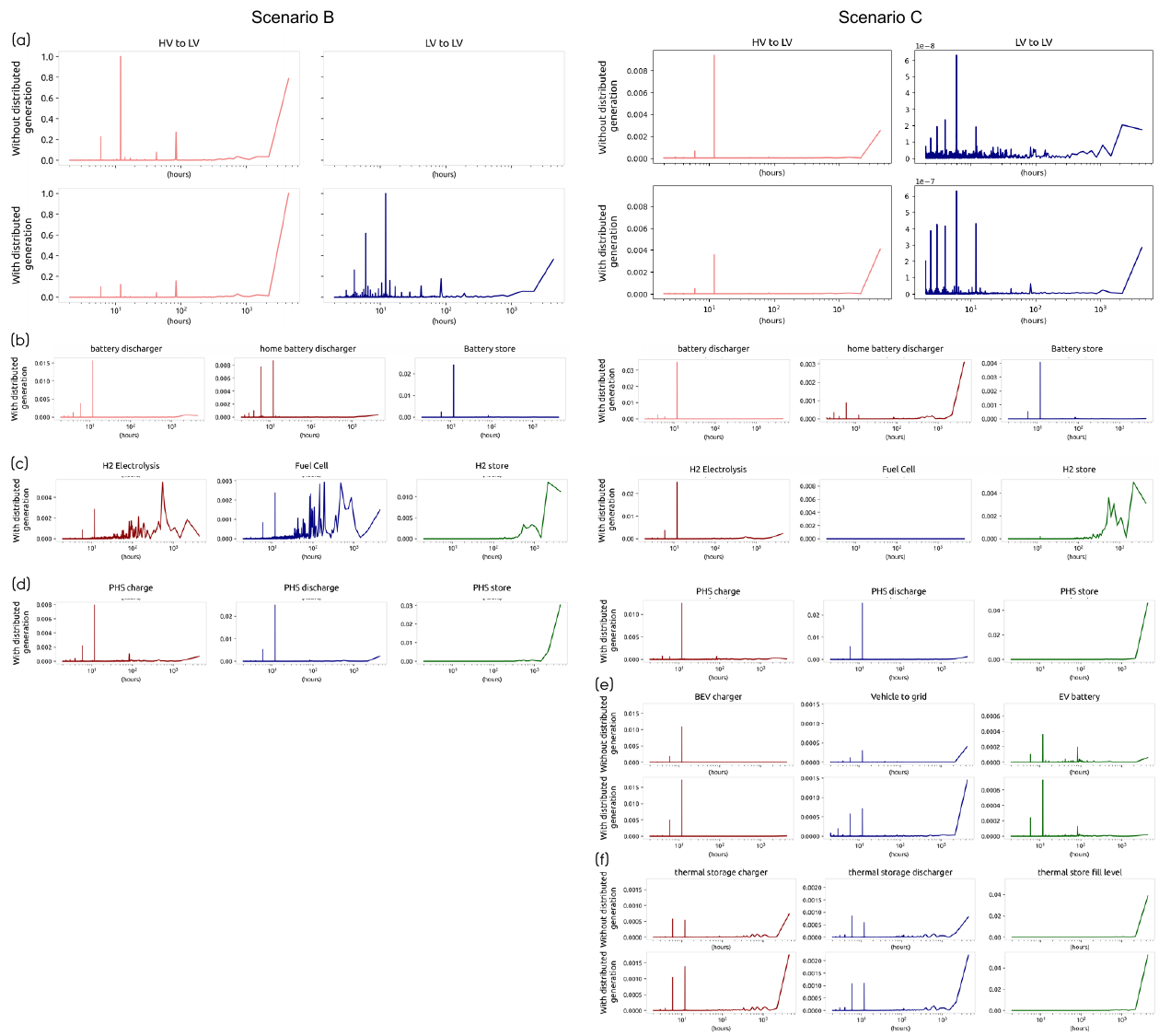}
\caption{Fourier power spectra of top) the distribution grid energy transfer time series from high-voltage (HV) to low-voltage (LV) and vice versa for scenario B and scenario C. The HV to LV graphs show three dominant cycles : (1) 6 hours, which is the average time solar power is available, (2) 12 hours, which shows the daily cycle of demand during day and night, and (3) 90 hours. The half-day, daily, and 3-day cycle can still be seen in the LV to HV (lower right) figure. Note: using a 1 year time series limits our ability to see the seasonal peak. Bottom figures show Fourier power spectra of time series for b) battery charger, battery discharger, and battery store hourly fill levels; c) pumped hydro storage (PHS) charger, PHS discharger, and PHS store hourly fill levels; d) battery electric vehicle (BEV) charger, vehicle to grid (V2G), and electric vehicle store hourly fill levels; and e) thermal storage charger, thermal storage discharger, and thermal store hourly fill levels.  
}
\end{figure}

\subsection*{S12. Role of distributed storage in the system}

\begin{figure}[H]
\renewcommand*{\thefigure}{S\arabic{figure}}

\includegraphics[width=0.8\textwidth,]{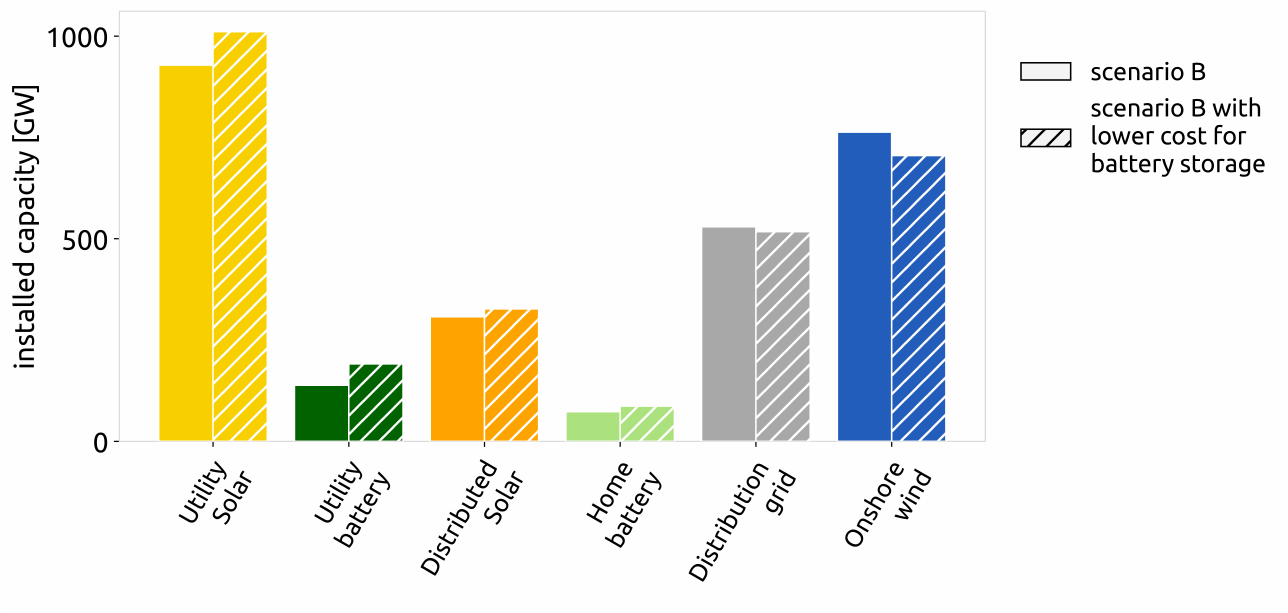}
\caption{Sensitivity of installed capacity of solar utility, utility batteries, distributed solar, home batteries, the distribution grid, and onshore wind for scenario B to battery storage costs. The costs for battery inverter (160 \texteuro /KW for utility battery  \protect\citeS{dea_table} and 228 \texteuro /KW for home battery  \protect\citeS{dea_table,ram2018global}) and lithium ion battery (142 \texteuro /KW for utility battery  \protect\citeS{dea_table} and 203 \texteuro /KW for home battery  \protect\citeS{dea_table,ram2018global}) assumed here are consistent with most studies  \protect\citeS{mauler2021battery}. However, considering the importance of their role in increasing the cost-efficiency of solar PV, a sensitivity case with utility battery inverter cost of 100 \texteuro /KW  \protect\citeS{wraalsen2022multiple} and utility battery cost of 100 \texteuro /KW \protect\citeS{mauler2021battery} is shown here. Home battery costs, lowered with the same ratio as utility battery costs compared to default assumption, are 160 \texteuro /KW for inverter and 127 \texteuro /KW for battery. When battery costs are lower, there is respectively a 8.8\% and 6.3\% increase in utility solar installed capacity and distributed solar installed capacity. Cost reduction as a result of including distributed solar in the system goes from 1.42\% (scenario B) to 1.84\% with cheaper battery storage. }
\end{figure}

\begin{figure}[H]
\renewcommand*{\thefigure}{S\arabic{figure}}

\includegraphics[width=0.83\textwidth,]{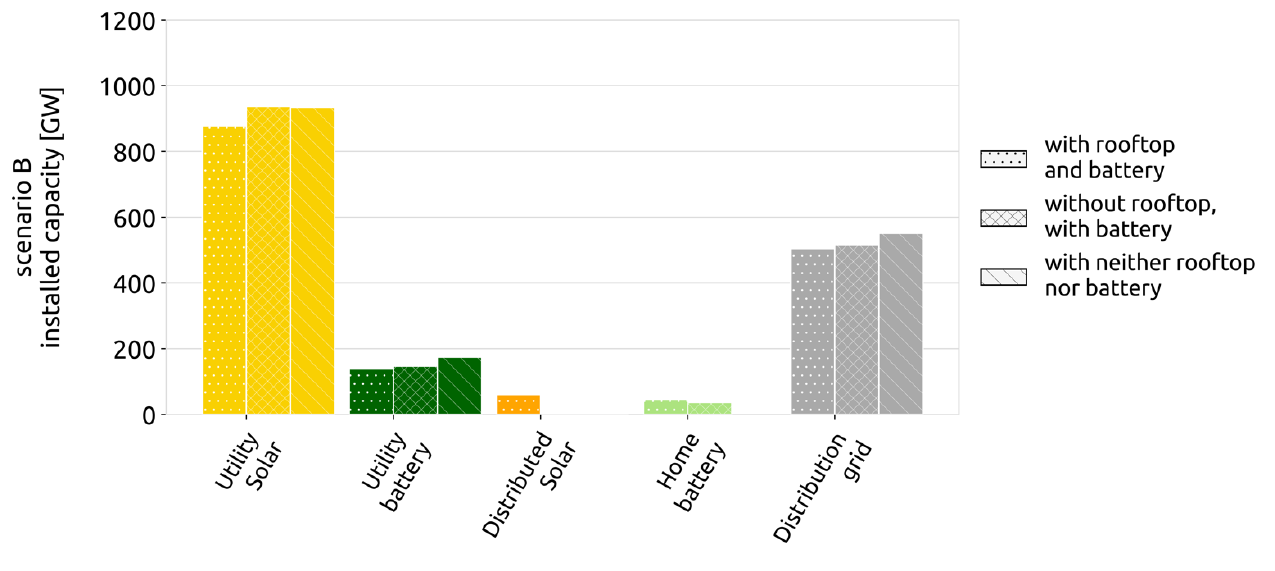}
\caption{Installed capacity of solar utility, utility batteries, distributed solar, home batteries, and the distribution grid for scenario B with/without home batteries and distributed solar. The figure shows that the system installs home batteries even without the presence of distributed solar.}
\end{figure}

\begin{figure}[H]
\renewcommand*{\thefigure}{S\arabic{figure}}

\includegraphics[width=0.9\textwidth, ]{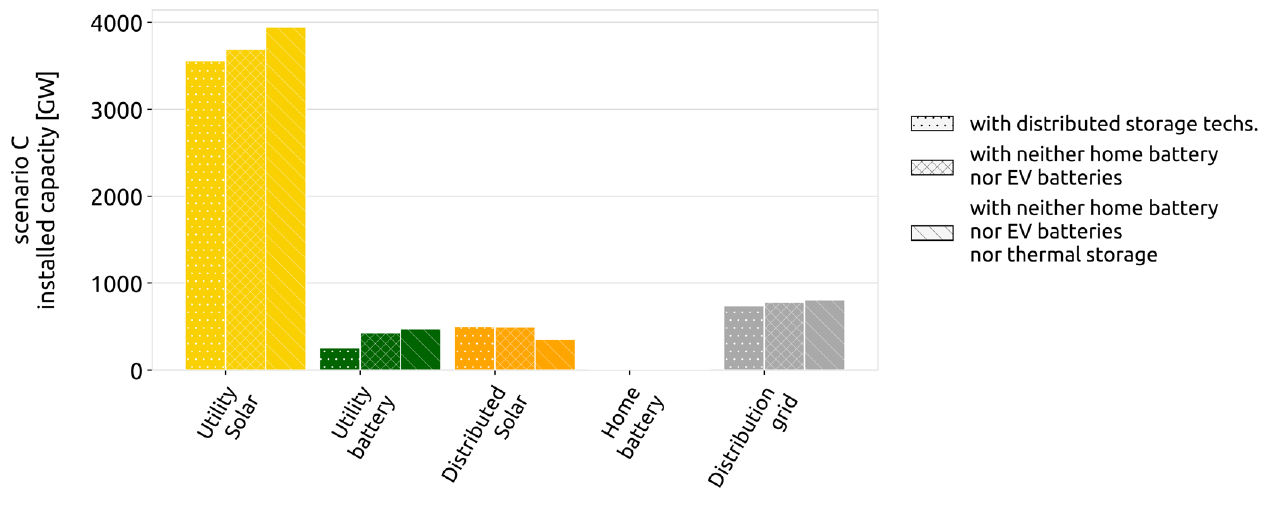}
\caption{Installed capacity of solar utility, utility batteries, distributed solar, home batteries, and the distribution grid for scenario C with and without the presence of different distributed storage technologies. There is no major difference observed without the presence of home batteries, and a small reduction when electric vehicle (EV) batteries and distributed thermal storage are also not included in the system. }
\end{figure}

\begin{figure}[H]
\renewcommand*{\thefigure}{S\arabic{figure}}

\includegraphics[width=0.9\textwidth, ]{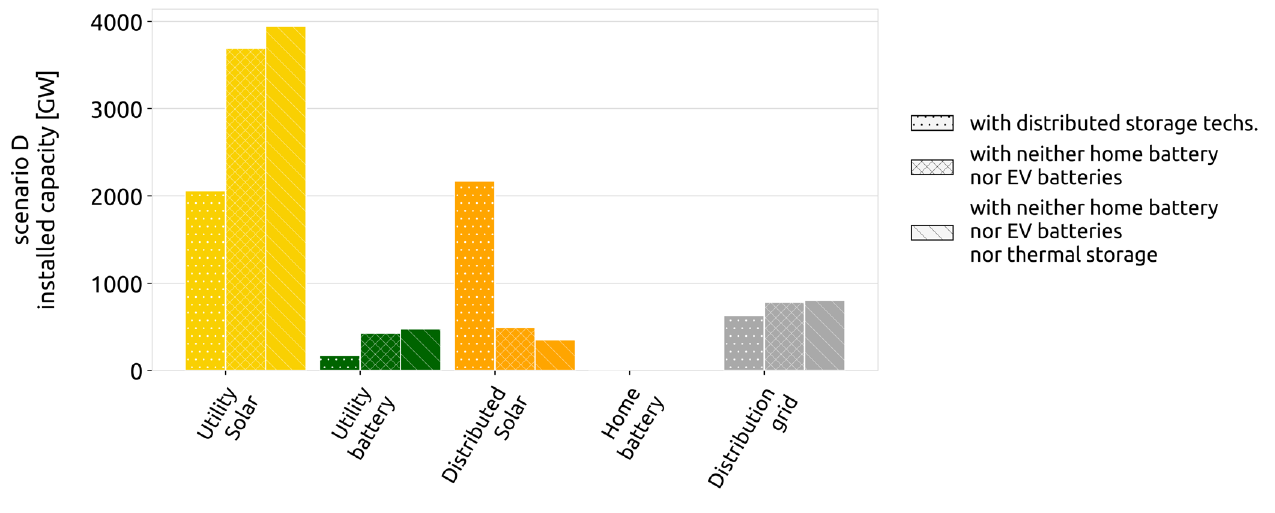}
\caption{Installed capacity of solar utility, utility batteries, distributed solar, home batteries, and the distribution grid for scenario D with and without the presence of different distributed storage technologies. There is a 80\% reduction in installed capacity of distributed solar when EV batteries are not included. Overall, There is a significant reduction of about 85\% in installed capacity of distributed solar when both EV batteries and thermal storage are not included in the system. This shows that the balancing provided by distributed storage majorly boosts the profitability of distributed solar in the system.}
\end{figure}

\begin{figure}[H]
\renewcommand*{\thefigure}{S\arabic{figure}}

\includegraphics[width=0.9\textwidth, center]{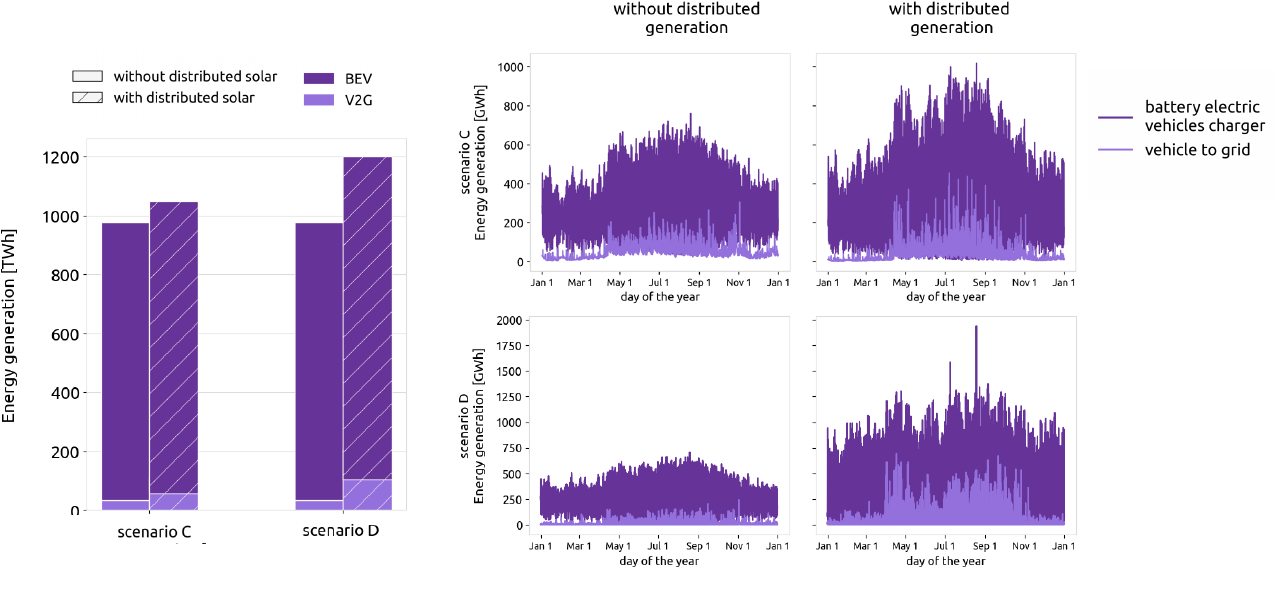}
\caption{Changes in the operation of electric vehicles battery's (BEV) charging and vehicle-to-grid (V2G) energy transfer for scenario C with and without distributed solar assuming default or high distributed solar potential. There is a clear rise in the energy transfer for both technologies in summer as they help balance distributed PV, especially when more solar is available. The high installed capacity of distributed PV for scenario D with high distributed solar potential means that the usage is increased for both BEV charging and V2G to help balance the large PV production at the low-voltage level of the grid. }
\end{figure}

\subsection*{S13. Sensitivity analysis for sector-coupled scenario}

\begin{figure}[H]
\renewcommand*{\thefigure}{S\arabic{figure}}
\includegraphics[width=0.8\textwidth,center]{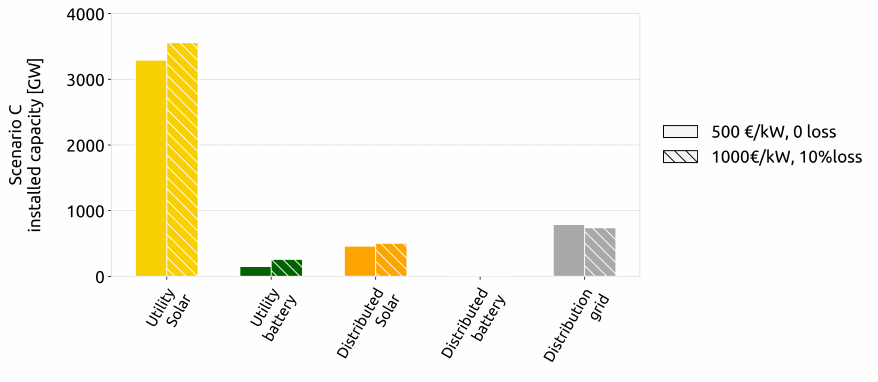}
\caption{Sensitivity of installed distributed solar to distribution grid cost and distribution grid losses when default distributed PV potential (504 GW) is assumed. There is only a 8.6\% reduction in the installed capacity of distributed solar. This shows that 500 GW of distributed solar is still cost-efficient for the system with lower grid costs and no grid losses.  }
\end{figure}

\begin{figure}[H]
\renewcommand*{\thefigure}{S\arabic{figure}}
\includegraphics[width=0.8\textwidth,center]{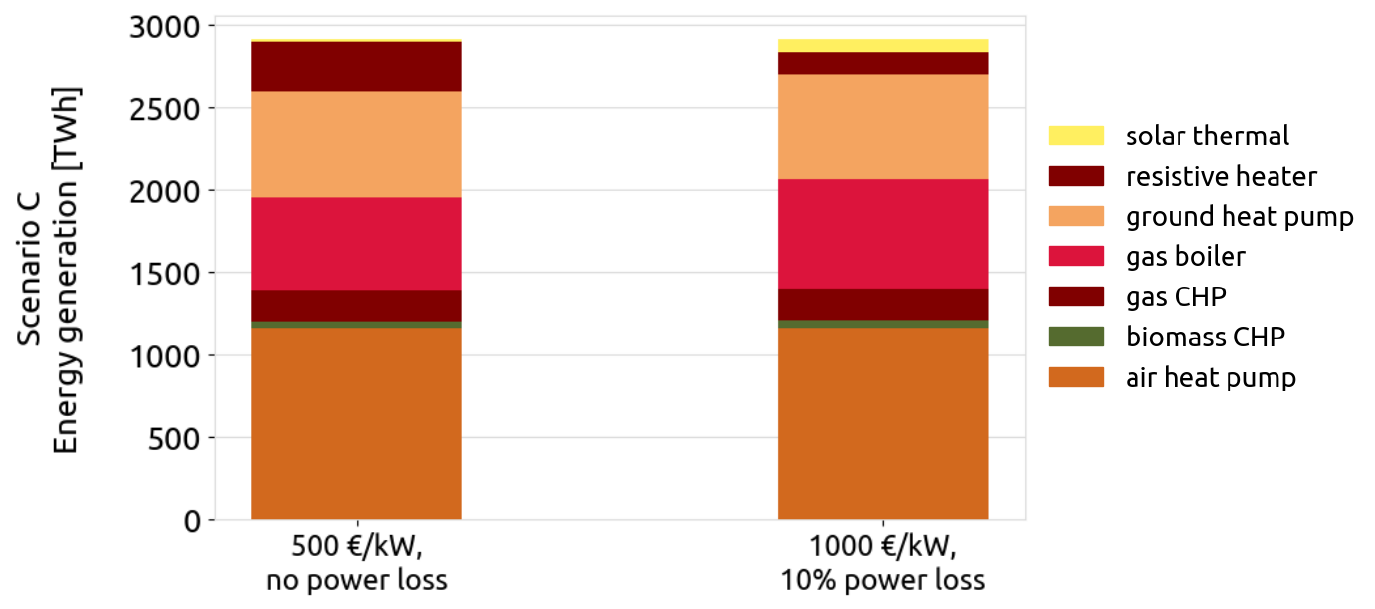}
\caption{Changes in heat generation mix for scenario C when distribution grid cost is 500 \texteuro /kW and power losses are 0\%. As mentioned in the paper, solar thermal is not selected in this case due to the fact that combined heat and power plants are more cost-efficient. }
\end{figure}

\begin{figure}[H]
\renewcommand*{\thefigure}{S\arabic{figure}}
\includegraphics[width=0.7\textwidth, center]{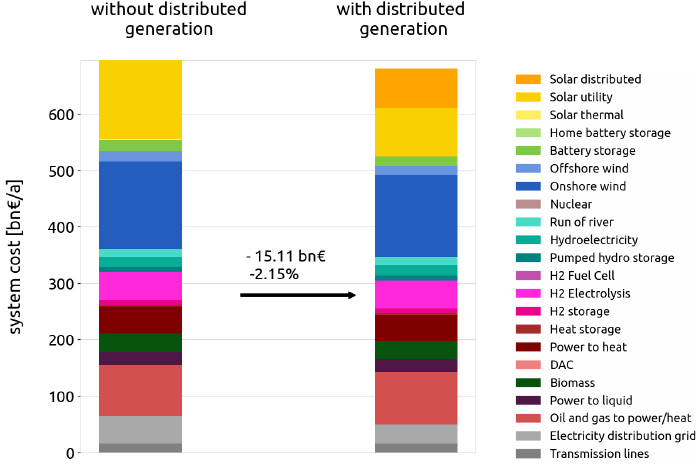}
\caption{Comparison of system costs for sector-coupled scenario with high distributed PV potential (3000 GW) with and without distributed technologies where distribution grid losses are 7\% and distribution grid cost is 500 \texteuro /KW. Cost savings (2.1\%) are less significant compared to scenario D (3.7\% savings) where assumptions for the distribution grid are 10\% losses and 1000 \texteuro /KW costs. However, the system still installs 1900 GW of distributed PV, which is 43\% of the total installed solar capacity equal to 4400 GW. This shows that even with more conservative assumptions regarding distribution grid, distributed solar is still profitable for the system. }
\end{figure}

\subsection*{S14. Cost assumptions}
Costs for all technologies and the source for each data is available at Github repository of PyPSA Technology Data \citeS{pypsacosts}.

\fontsize{7}{0.1}\selectfont
\renewcommand*{\thetable}{S\arabic{table}}
\setcounter{table}{2}
\begin{longtblr}[
 caption = {Projected cost assumptions for major technologies in 2030.},
  label = none,
  entry = none,
]{
  rowhead=1,
  width = \linewidth,
  colspec = {Q[304]Q[240]Q[102]Q[292]},
  cell{2}{1} = {r=3}{},
  cell{5}{1} = {r=3}{},
  cell{8}{1} = {r=3}{},
  cell{11}{1} = {r=3}{},
  cell{14}{1} = {r=5}{},
  cell{19}{1} = {r=6}{},
  cell{25}{1} = {r=4}{},
  cell{29}{1} = {r=4}{},
  cell{33}{1} = {r=2}{},
  cell{35}{1} = {r=5}{},
  cell{40}{1} = {r=8}{},
  cell{48}{1} = {r=5}{},
  cell{53}{1} = {r=5}{},
  cell{58}{1} = {r=8}{},
  cell{66}{1} = {r=6}{},
  cell{72}{1} = {r=5}{},
  cell{77}{1} = {r=5}{},
  cell{82}{1} = {r=5}{},
  cell{87}{1} = {r=3}{},
  cell{90}{1} = {r=9}{},
  cell{99}{1} = {r=8}{},
  cell{107}{1} = {r=6}{},
  cell{113}{1} = {r=7}{},
  cell{120}{1} = {r=4}{},
  cell{124}{1} = {r=5}{},
  cell{129}{1} = {r=5}{},
  cell{134}{1} = {r=2}{},
  cell{136}{1} = {r=5}{},
  cell{141}{1} = {r=4}{},
  cell{145}{1} = {r=5}{},
  cell{150}{1} = {r=4}{},
  cell{154}{1} = {r=4}{},
  cell{158}{1} = {r=8}{},
  cell{166}{1} = {r=6}{},
  cell{172}{1} = {r=5}{},
  cell{177}{1} = {r=5}{},
  cell{182}{1} = {r=2}{},
  cell{184}{1} = {r=4}{},
  cell{188}{1} = {r=2}{},
  cell{190}{1} = {r=4}{},
  cell{194}{1} = {r=6}{},
  cell{200}{1} = {r=4}{},
  cell{204}{1} = {r=4}{},
  cell{208}{1} = {r=7}{},
  cell{215}{1} = {r=4}{},
  cell{219}{1} = {r=4}{},
  cell{223}{1} = {r=3}{},
  hline{1} = {-}{0.08em},
  hline{2,5,8,11,14,19,25,29,33,35,40,48,53,58,66,72,77,82,87,90,99,107,113,120,124,129,134,136,141,145,150,154,158,166,172,177,182,184,188,190,194,200,204,208,215,219,223,226-228} = {-}{},
}
Technology                         & Parameter                     & Value     & Unit                              \\
HVAC overhead                      & FOM                           & 2         & \%/year                           \\
                                   & investment                    & 432.97    & EUR/MW/km                         \\
                                   & lifetime                      & 40        & years                             \\
HVDC inverter pair                 & FOM                           & 2         & \%/year                           \\
                                   & investment                    & 162364.82 & EUR/MW                            \\
                                   & lifetime                      & 40        & years                             \\
HVDC overhead                      & FOM                           & 2         & \%/year                           \\
                                   & investment                    & 432.97    & EUR/MW/km                         \\
                                   & lifetime                      & 40        & years                             \\
HVDC submarine                     & FOM                           & 0.35      & \%/year                           \\
                                   & investment                    & 471.16    & EUR/MW/km                         \\
                                   & lifetime                      & 40        & years                             \\
OCGT                               & FOM                           & 1.78      & \%/year                           \\
                                   & VOM                           & 4.5       & EUR/MWh                           \\
                                   & efficiency                    & 0.41      & per unit                          \\
                                   & investment                    & 435.24    & EUR/kW                            \\
                                   & lifetime                      & 25        & years                             \\
CCGT                               & VOM                           & 4.2       & EUR/MWh                           \\
                                   & c\_b                          & 2         & 50\degree C/100\degree C                        \\
                                   & c\_v                          & 0.15      & 50\degree C/100\degree C                        \\
                                   & efficiency                    & 0.58      & per unit                          \\
                                   & investment                    & 830       & EUR/kW                            \\
                                   & lifetime                      & 25        & years                             \\
PHS                                & FOM                           & 1         & \%/year                           \\
                                   & efficiency                    & 0.75      & per unit                          \\
                                   & investment                    & 2208.16   & EUR/kWel                          \\
                                   & lifetime                      & 80        & years                             \\
battery inverter                   & FOM                           & 0.34      & \%/year                           \\
                                   & efficiency                    & 0.96      & per unit                          \\
                                   & investment                    & 160       & EUR/kW                            \\
                                   & lifetime                      & 10        & years                             \\
battery storage                    & investment                    & 142       & EUR/kWh                           \\
                                   & lifetime                      & 25        & years                             \\
biomass                            & FOM                           & 4.53      & \%/year                           \\
                                   & efficiency                    & 0.47      & per unit                          \\
                                   & fuel                          & 7         & EUR/MWhth                         \\
                                   & investment                    & 2209      & EUR/kWel                          \\
                                   & lifetime                      & 30        & years                             \\
biomass CHP                        & FOM                           & 3.58      & \%/year                           \\
                                   & VOM                           & 2.1       & EUR/MWh\_e                        \\
                                   & c\_b                          & 0.46      & 40\degree C/80\degree C                       \\
                                   & c\_v                          & 1         & 40\degree C/80\degree C                       \\
                                   & efficiency                    & 0.3       & per unit                          \\
                                   & efficiency-heat               & 0.71      & per unit                          \\
                                   & investment                    & 3210.28   & EUR/kW\_e                         \\
                                   & lifetime                      & 25        & years                             \\
biomass boiler                     & FOM                           & 7.49      & \%/year                           \\
                                   & efficiency                    & 0.86      & per unit                          \\
                                   & investment                    & 649.3     & EUR/kW\_th                        \\
                                   & lifetime                      & 20        & years                             \\
                                   & pelletizing cost              & 9         & EUR/MWh\_pellets                  \\
central air-sourced heat pump      & FOM                           & 0.23      & \%/year                           \\
                                   & VOM                           & 2.51      & EUR/MWh\_th                       \\
                                   & efficiency                    & 3.6       & per unit                          \\
                                   & investment                    & 856.25    & EUR/kW\_th                        \\
                                   & lifetime                      & 25        & years                             \\
central gas CHP                    & FOM                           & 3.32      & \%/year                           \\
                                   & VOM                           & 4.2       & EUR/MWh                           \\
                                   & c\_b                          & 1         & 50\degree C/100\degree C                        \\
                                   & c\_v                          & 0.17      & per unit                          \\
                                   & efficiency                    & 0.41      & per unit                          \\
                                   & investment                    & 560       & EUR/kW                            \\
                                   & lifetime                      & 25        & years                             \\
                                   & p\_nom\_ratio                 & 1         & per unit                          \\
central gas CHP CC                 & FOM                           & 3.32      & \%/year                           \\
                                   & VOM                           & 4.2       & EUR/MWh                           \\
                                   & c\_b                          & 1         & 50\degree C/100\degree C                        \\
                                   & efficiency                    & 0.41      & per unit                          \\
                                   & investment                    & 560       & EUR/kW                            \\
                                   & lifetime                      & 25        & years                             \\
central gas boiler                 & FOM                           & 3.8       & \%/year                           \\
                                   & VOM                           & 1         & EUR/MWh\_th                       \\
                                   & efficiency                    & 1.04      & per unit                          \\
                                   & investment                    & 50        & EUR/kW\_th                        \\
                                   & lifetime                      & 25        & years                             \\
central ground-sourced heat pump   & FOM                           & 0.39      & \%/year                           \\
                                   & VOM                           & 1.25      & EUR/MWh\_th                       \\
                                   & efficiency                    & 1.73      & per unit                          \\
                                   & investment                    & 507.6     & EUR/kW\_th excluding drive energy \\
                                   & lifetime                      & 25        & years                             \\
central resistive heater           & FOM                           & 1.7       & \%/year                           \\
                                   & VOM                           & 1         & EUR/MWh\_th                       \\
                                   & efficiency                    & 0.99      & per unit                          \\
                                   & investment                    & 60        & EUR/kW\_th                        \\
                                   & lifetime                      & 20        & years                             \\
central solar thermal              & FOM                           & 1.4       & \%/year                           \\
                                   & investment                    & 140000    & EUR/1000m2                        \\
                                   & lifetime                      & 20        & years                             \\
central solid biomass CHP          & FOM                           & 2.87      & \%/year                           \\
                                   & VOM                           & 4.58      & EUR/MWh\_e                        \\
                                   & c\_b                          & 0.35      & 50\degree C/100\degree C                      \\
                                   & c\_v                          & 1         & 50\degree C/100\degree C                      \\
                                   & efficiency                    & 0.27      & per unit                          \\
                                   & efficiency-heat               & 0.82      & per unit                          \\
                                   & investment                    & 3349.49   & EUR/kW\_e                         \\
                                   & lifetime                      & 25        & years                             \\
                                   & p\_nom\_ratio                 & 1         & per unit                          \\
central solid biomass CHP CC       & FOM                           & 2.87      & \%/year                           \\
                                   & VOM                           & 4.58      & EUR/MWh\_e                        \\
                                   & c\_b                          & 0.35      & 50\degree C/100\degree C                      \\
                                   & c\_v                          & 1         & 50\degree C/100\degree C                      \\
                                   & efficiency                    & 0.27      & per unit                          \\
                                   & efficiency-heat               & 0.82      & per unit                          \\
                                   & investment                    & 4921.02   & EUR/kW\_e                         \\
                                   & lifetime                      & 25        & years                             \\
central water tank storage         & FOM                           & 0.55      & \%/year                           \\
                                   & investment                    & 0.54      & EUR/kWhCapacity                   \\
                                   & lifetime                      & 25        & years                             \\
                                   & FOM                           & 2         & \%/year                           \\
                                   & investment                    & 67.63     & EUR/m\^{}3-H2O                    \\
                                   & lifetime                      & 30        & years                             \\
coal                               & CO2 intensity                 & 0.34      & tCO2/MWh\_th                      \\
                                   & FOM                           & 1.6       & \%/year                           \\
                                   & VOM                           & 3.5       & EUR/MWh\_e                        \\
                                   & efficiency                    & 0.33      & per unit                          \\
                                   & fuel                          & 8.15      & EUR/MWh\_th                       \\
                                   & investment                    & 3845.51   & EUR/kW\_e                         \\
                                   & lifetime                      & 40        & years                             \\
decentral CHP                      & FOM                           & 3         & \%/year                           \\
                                   & discount rate                 & 0.04      & per unit                          \\
                                   & investment                    & 1400      & EUR/kWel                          \\
                                   & lifetime                      & 25        & years                             \\
decentral air-sourced heat pump    & FOM                           & 3         & \%/year                           \\
                                   & discount rate                 & 0.04      & per unit                          \\
                                   & efficiency                    & 3.6       & per unit                          \\
                                   & investment                    & 850       & EUR/kW\_th                        \\
                                   & lifetime                      & 18        & years                             \\
decentral gas boiler               & FOM                           & 6.69      & \%/year                           \\
                                   & discount rate                 & 0.04      & per unit                          \\
                                   & efficiency                    & 0.98      & per unit                          \\
                                   & investment                    & 296.82    & EUR/kW\_th                        \\
                                   & lifetime                      & 20        & years                             \\
decentral gas boiler connection    & investment                    & 185.51    & EUR/kW\_th                        \\
                                   & lifetime                      & 50        & years                             \\
decentral ground-sourced heat pump & FOM                           & 1.82      & \%/year                           \\
                                   & discount rate                 & 0.04      & per unit                          \\
                                   & efficiency                    & 3.9       & per unit                          \\
                                   & investment                    & 1400      & EUR/kW\_th                        \\
                                   & lifetime                      & 20        & years                             \\
decentral oil boiler               & FOM                           & 2         & \%/year                           \\
                                   & efficiency                    & 0.9       & per unit                          \\
                                   & investment                    & 156.01    & EUR/kWth                          \\
                                   & lifetime                      & 20        & years                             \\
decentral resistive heater         & FOM                           & 2         & \%/year                           \\
                                   & discount rate                 & 0.04      & per unit                          \\
                                   & efficiency                    & 0.9       & per unit                          \\
                                   & investment                    & 100       & EUR/kWhth                         \\
                                   & lifetime                      & 20        & years                             \\
decentral solar thermal            & FOM                           & 1.3       & \%/year                           \\
                                   & discount rate                 & 0.04      & per unit                          \\
                                   & investment                    & 270000    & EUR/1000m2                        \\
                                   & lifetime                      & 20        & years                             \\
decentral water tank storage       & FOM                           & 1         & \%/year                           \\
                                   & discount rate                 & 0.04      & per unit                          \\
                                   & investment                    & 18.38     & EUR/kWh                           \\
                                   & lifetime                      & 20        & years                             \\
direct air capture                 & FOM                           & 4.95      & \%/year                           \\
                                   & compression-electricity-input & 0.15      & MWh/tCO2                          \\
                                   & compression-heat-output       & 0.2       & MWh/tCO2                          \\
                                   & electricity-input             & 0.4       & MWh\_el/t\_CO2                    \\
                                   & heat-input                    & 1.6       & MWh\_th/t\_CO2                    \\
                                   & heat-output                   & 1         & MWh/tCO2                          \\
                                   & investment                    & 6000000   & EUR/(tCO2/h)                      \\
                                   & lifetime                      & 20        & years                             \\
electricity distribution grid      & FOM                           & 2         & \%/year                           \\
                                   & investment                    & 500       & EUR/kW                            \\
                                   & lifetime                      & 40        & years                             \\
                                   & FOM                           & 2         & \%/year                           \\
                                   & investment                    & 140       & EUR/kW                            \\
                                   & lifetime                      & 40        & years                             \\
electrolysis                       & FOM                           & 2         & \%/year                           \\
                                   & efficiency                    & 0.8       & per unit                          \\
                                   & efficiency-heat               & 0.17      & per unit                          \\
                                   & investment                    & 407.58    & EUR/kW\_e                         \\
                                   & lifetime                      & 30        & years                             \\
fuel cell                          & FOM                           & 5         & \%/year                           \\
                                   & c\_b                          & 1.25      & 50\degree C/100\degree C                        \\
                                   & efficiency                    & 0.58      & per unit                          \\
                                   & investment                    & 1100      & EUR/kW\_e                         \\
                                   & lifetime                      & 10        & years                             \\
gas                                & CO2 intensity                 & 0.2       & tCO2/MWh\_th                      \\
                                   & fuel                          & 20.1      & EUR/MWh\_th                       \\
home battery inverter              & FOM                           & 0.34      & \%/year                           \\
                                   & efficiency                    & 0.96      & per unit                          \\
                                   & investment                    & 228.06    & EUR/kW                            \\
                                   & lifetime                      & 10        & years                             \\
home battery storage               & investment                    & 202.9     & EUR/kWh                           \\
                                   & lifetime                      & 25        & years                             \\
hydro                              & FOM                           & 1         & \%/year                           \\
                                   & efficiency                    & 0.9       & per unit                          \\
                                   & investment                    & 2208.16   & EUR/kWel                          \\
                                   & lifetime                      & 80        & years                             \\
nuclear                            & FOM                           & 1.4       & \%/year                           \\
                                   & VOM                           & 3.5       & EUR/MWh\_e                        \\
                                   & efficiency                    & 0.33      & per unit                          \\
                                   & fuel                          & 2.6       & EUR/MWh\_th                       \\
                                   & investment                    & 7940.45   & EUR/kW\_e                         \\
                                   & lifetime                      & 40        & years                             \\
onshore wind                       & FOM                           & 1.22      & \%/year                           \\
                                   & VOM                           & 1.35      & EUR/MWh                           \\
                                   & investment                    & 1035.56   & EUR/kW                            \\
                                   & lifetime                      & 30        & years                             \\
offshore wind                      & FOM                           & 2.32      & \%/year                           \\
                                   & VOM                           & 0.02      & EUR/MWhel                         \\
                                   & investment                    & 1523.55   & EUR/kW\_e, 2020                   \\
                                   & lifetime                      & 30        & years                             \\
oil                                & CO2 intensity                 & 0.26      & tCO2/MWh\_th                      \\
                                   & FOM                           & 2.46      & \%/year                           \\
                                   & VOM                           & 6         & EUR/MWh                           \\
                                   & efficiency                    & 0.35      & per unit                          \\
                                   & fuel                          & 50        & EUR/MWhth                         \\
                                   & investment                    & 343       & EUR/kW                            \\
                                   & lifetime                      & 25        & years                             \\
ror                                & FOM                           & 2         & \%/year                           \\
                                   & efficiency                    & 0.9       & per unit                          \\
                                   & investment                    & 3312.24   & EUR/kWel                          \\
                                   & lifetime                      & 80        & years                             \\
solar-rooftop                      & FOM                           & 1.42      & \%/year                           \\
                                   & discount rate                 & 0.04      & per unit                          \\
                                   & investment                    & 636.66    & EUR/kW\_e                         \\
                                   & lifetime                      & 40        & years                             \\
solar-utility                      & FOM                           & 2.48      & \%/year                           \\
                                   & investment                    & 347.56    & EUR/kW\_e                         \\
                                   & lifetime                      & 40        & years                             \\
water tank charger                 & efficiency                    & 0.84      & per unit                          \\
water tank discharger              & efficiency                    & 0.84      & per unit                          
\end{longtblr}

\clearpage